\documentclass[5p,times,twocolumn]{elsarticle}

\usepackage{lineno}
\modulolinenumbers[5]
\usepackage{amsmath}
\usepackage{amsfonts,tabularx}
\usepackage{algorithm}
\usepackage[noend]{algpseudocode}

\usepackage{amsmath,amssymb,bm,graphicx,epstopdf}
\usepackage{amsthm}
\allowdisplaybreaks
\usepackage{dblfloatfix}
\usepackage{soul}
\usepackage[dvipsnames]{xcolor}

\usepackage{multirow}
\usepackage{graphicx}
\usepackage{graphbox}
\usepackage{mathtools}
\usepackage[colorlinks=true]{hyperref}
\hypersetup{
    colorlinks,
    citecolor=blue,
    linkcolor=blue
}
\usepackage[caption=false, font=footnotesize]{subfig}
\usepackage{ctable}

\usepackage{framed} 
\usepackage{multicol} 

\usepackage{nomencl} 
\makenomenclature
\usepackage{etoolbox}
\renewcommand\nomgroup[1]{%
  \item[\bfseries
  \ifstrequal{#1}{A}{Controller Abbreviations}{%
  \ifstrequal{#1}{I}{Indices and Related Constants}{}%
  \ifstrequal{#1}{V}{Variables}{}}
]}

\usepackage[noend]{algpseudocode}

\makeatletter
\newcommand{\multiline}[1]{%
  \begin{tabularx}{\dimexpr\linewidth-\ALG@thistlm}[t]{@{}X@{}}
    #1
  \end{tabularx}
}
\makeatother

\usepackage{amsmath,amsfonts,bm}

\newcommand{\cvxpylayer}{MPC-CL}
\newcommand{\dpc}{DPC}
\newcommand{\rl}{RLC}

\renewcommand{\u}{\mathbf{u}}
\newcommand{\x}{\mathbf{x}}
\newcommand{\s}{\mathbf{s}}
\newcommand{\w}{\mathbf{w}}

\newcommand{\param}{\bm{\theta}}

\begin{document}

\begin{frontmatter}


\title{From Model-Based to Model-Free: Learning Building Control for Demand Response}

\author[nreladdress,maplewelladdress]{David Biagioni}
\ead{dave@maplewelleng.com}

\author[nreladdress]{Xiangyu Zhang \corref{correspondingauthor}}
\cortext[correspondingauthor]{Corresponding author}
\ead{Xiangyu.Zhang@nrel.gov }

\author[nreladdress,stanfordaddress]{Christiane Adcock}
\ead{janiad@stanford.edu}

\author[nreladdress]{Michael Sinner}
\ead{Michael.Sinner@nrel.gov}

\author[nreladdress]{Peter Graf}
\ead{Peter.Graf@nrel.gov}

\author[nreladdress]{Jennifer King}
\ead{Jennifer.King@nrel.gov}

\address[nreladdress]{National Renewable Energy Laboratory, Golden, CO 80401, U.S.A.}
\address[stanfordaddress]{Stanford University, Stanford, CA 94305, U.S.A.}
\address[maplewelladdress]{Maplewell Energy, Broomfield, CO 80021, U.S.A.}

\begin{abstract}
Grid-interactive building control is a challenging and important problem for reducing carbon emissions, increasing energy efficiency, and supporting the electric power grid. Currently researchers and practitioners are confronted with a choice of control strategies ranging from model-free (purely data-driven) to model-based (directly incorporating physical knowledge) to hybrid methods that combine data and models. In this work, we identify state-of-the-art methods that span this methodological spectrum and evaluate their performance for multi-zone building HVAC control in the context of three demand response programs. We demonstrate, in this context, that hybrid methods offer many benefits over both purely model-free and model-based methods as long as certain requirements are met. In particular, hybrid controllers are relatively sample efficient, fast online, and high accuracy so long as the test case falls within the distribution of training data. Like all data-driven methods, hybrid controllers are still subject to generalization errors when applied to out-of-sample scenarios. Key takeaways for control strategies are summarized and the developed software framework is open-sourced.
\end{abstract}


\end{frontmatter}

\section{Introduction}\label{intro}

The increasingly frequent occurrence of extreme weather events and their severe social and economic consequences make climate change one of the most pressing problems currently faced by human society \cite{stott2016climate}. To address this, many countries have proposed quantitative targets to reverse the outcome caused by human-induced climate change and guide the transition to a decarbonized society. For example, the U.S. government aims to decarbonize the power sector by 2035 and reach net zero emissions by 2050 \cite{room2021fact}. Similarly, China’s ``30-60'' plan targets to hit “carbon peak” and “carbon neutral” by 2030 and 2060, respectively \cite{downie2021getting}. A key part of reaching these goals will be reducing building energy consumption, which in the U.S., for example, accounts for 70\% of national electricity use \cite{somasundaram2014reference} and 30\% of national carbon emissions \cite{leung2018decarbonizing}. An effective set of tools to reduce building energy consumption are technologies for grid-interactive efficient buildings (GEBs.) According to \cite{satchwell2021national}, adopting these technologies across the U.S. would reduce power sector emissions by 6\% by 2030 and save \$100-200 billion by 2040. To achieve these reductions, it is urgent and necessary to develop technologies that enable buildings to be grid-interactive, such as through demand response (DR) programs. 

One successful method for building-grid interaction through DR is model predictive control (MPC). MPC has been widely studied for optimizing building-side control objectives \cite{privara2011model, kim2020model}; see \cite{drgona2020all} for a comprehensive review. Many recent works add grid service objectives to the MPC formulation to deliver control that balances the trade-offs between thermal comfort, energy cost, and DR requirements. For example, Tang et al. \cite{tang2019model} propose leveraging thermal energy storage in buildings to provide fast DR while maintaining indoor comfort and Hu et al. \cite{hu2019price} design a mixed integer linear programming-based MPC controller to make building floor heating systems grid responsive. However, as pointed out by \cite{drgona2020all}, while MPC controllers have been researched for more than a decade, they have not yet been widely adopted in buildings owing to challenges in creating accurate yet simple predictive models and solving potentially compute-intensive optimization problems during online control.

To circumvent the difficulties with MPC, researchers have tried purely data-driven methods, such as model-free reinforcement learning (RL). Building simulators such as EnergyPlus or thermal dynamics models learned from smart thermostat data can be used for RL controller training, as demonstrated in \cite{wei2017deep, zhang2019whole, zhang2020edge, zhang2022two}. As RL has less stringent requirements for building models than MPC, it is easier to obtain models for new buildings, although several reviews highlight the need for more research on transfer learning to further improve generalization \cite{schmidt2018, wang2020}. RL also has the benefit that online operation only requires a single forward pass through a policy, such as a neural network, which is generally much faster than the online optimization required by MPC. For a more in-depth review on using RL for DR, refer to \cite{vazquez2019reinforcement}. The use of model-free RL comes at a price, however: training can require large amounts of data and time, and it can be challenging to enforce safety or physical constraints. The primary approach for the later has been adding soft penalties for constraint violation, which only encourages but doesn't strictly enforce the constraints. While recent research has attempted to address some of these challenges for building control using transfer learning \cite{xu2020one, zhang2020transferable, pinto2022transfer} and safe RL \cite{yu2021district}, it is still an active area of research.

Recently, hybrids of MPC and model-free RL have shown promise for building control. Intuitively, a model-based method such as MPC embodies physical domain knowledge and thus can guide a data-driven learning process while a learning-based method such as RL can adapt the model to a specific system. Based on this intuition, hybrid control methods should train offline faster than RL, run online faster than MPC, and identify at least as good control actions as either method. Several hybrid methods have been proposed in recent years, notably MPC with a learned terminal cost (RL-MPC) \cite{kowli2012coordinating,arroyo2022reinforced}, approximate MPC (aMPC) based on imitation learning \cite{drgona2018approximate, karg2020efficient}, and differentiable predictive control (DPC) \cite{drgona2020learning} based on differentiable programming \cite{jin2020pontryagin, amos2018differentiable}. aMPC and DPC use a similar approach and have comparable control performance, but DPC is simpler to implement and has lower training cost \cite{drgona2020learning}.

These works include some comparisons of hybrid methods to MPC and/or RL for building control. For single building control, \cite{arroyo2022reinforced} compares RL-MPC to MPC and RL and \cite{drgona2020learning} compares DPC to MPC and RL. For multi-zone building control, \cite{drgona2018approximate} compares aMPC to MPC. These studies each compare one hybrid controller to one or two baseline controllers rather than comparing hybrid controllers to each other and they make these comparisons on a limited set of metrics, such as only control performance and/or online evaluation time. Also, they only consider baseline building energy control rather than grid-interactive control. In addition, some of the hybrid methods are applied to building control in a limited capacity: RL-MPC either uses a restrictive form for the terminal cost \cite{kowli2012coordinating} or uses a computationally expensive sampling method \cite{arroyo2022reinforced} and DPC has only been trained using a linearized building model \cite{drgona2020learning}. Finally, the above comparisons use a mix of building models and learning frameworks, some of which are not released open-source, making it challenging to reproduce, build upon, or compare between these works. 

\begin{figure}[]
\centering
\includegraphics[width=0.7\linewidth]{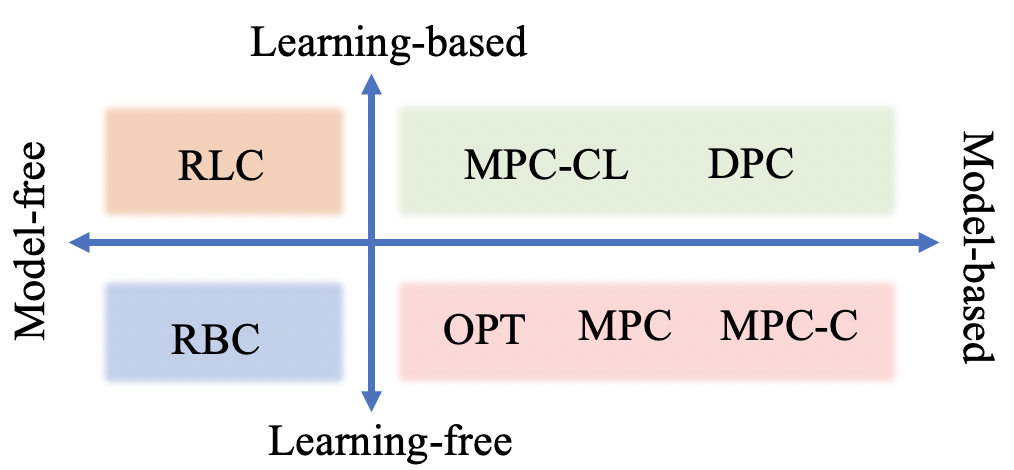}
\caption{A quadrant illustration on control methods studied in this paper based on two feature axes: (1) use of learning and (2) explicit use of a model. See the nomenclature section for controllers' full names and Section \ref{control} for detailed formulations.}
\label{fig-controller-quadrant}
\end{figure}

To address the knowledge gaps described above, we present the first paper providing a comprehensive comparison of model-based, learning-based, and hybrid methods for building control. The novel contributions of this study are that we:

1. Compare a full spectrum of state-of-the-art control methods from model-based to model-free and from learning-based to learning-free, as shown in Figure. \ref{fig-controller-quadrant}.

2. Evaluate control methods on a thorough set of metrics: training performance, control performance, generalizability, and online computational time. Based on our experiments, we summarize the pros and cons of each method and for what test cases a given method is suitable.

3. Include DR programs in the analysis of building controllers and extend these methods to grid-interactive control.

4. Use robust, general forms of hybrid methods to ensure a fair comparison between methods. Unlike previous works, we apply RL-MPC using a general form for the cost function and a computationally efficient training approach and we train DPC on a nonlinear building model. 

5. Implement seven control methods, two building models, and three DR programs in a modular, documented, open-source code base \cite{lbc_code} (URL: \url{https://github.com/NREL/learning-building-control}) to enable future comparisons to and extensions of this work.

The rest of the paper is arranged as follows: Section \ref{problem-ms} covers the problem formulation by describing our building model, cost function, DR programs, general building control problem, and measured signals. Section \ref{control} explains the control methods compared in this work. Training methods for the data-driven methods are described in Section \ref{training}. Experimental details and results are presented in Section \ref{results} and conclusions in Section \ref{sec-conclusion}.

\newcommand{\p}[1]{P^\text{#1}}
\newcommand{\pchill}{\p{chiller}}
\newcommand{\pflow}{\p{fan}}
\newcommand{\ptotal}{\p{total}}
\newcommand{\COP}{\text{COP}}
\newcommand{\Tout}{T^{\text{out}}}
\newcommand{\Tsupply}{T^{\text{supply}}}
\newcommand{\Tzones}{\mathbf{T}}

\begin{table}[!t]
\footnotesize
\label{tab:nomenclature}
  \begin{framed}
    \printnomenclature
  \end{framed}
\end{table}

\nomenclature[A1]{OPT}{Optimal Open-Loop Control}
\nomenclature[A2]{MPC}{Nonconvex model predictive control}
\nomenclature[A3]{MPC-C}{Convex model predictive control}
\nomenclature[A4]{MPC-CL}{Convex model predictive control with learned terminal cost}
\nomenclature[A5]{DPC}{Differentiable predictive control}
\nomenclature[A6]{RLC}{Reinforcement learning control}
\nomenclature[A7]{RBC}{Rule based control}

\nomenclature[I]{$t; N$}{Control step index; total control steps}
\nomenclature[I]{$\tau$}{Duration of control step}
\nomenclature[I]{$k; K$}{Lookahead step index; total lookahead steps}
\nomenclature[I]{$z$}{Number of thermal zones in the building}

\nomenclature[V]{$\x$}{State variable in building dynamics}
\nomenclature[V]{$\u$}{Control action}
\nomenclature[V]{$\w$}{Exogenous inputs}

\section{Model and Control Problem Formulation} \label{problem-ms}

To investigate building control under demand response (DR), we focus on controlling the heating, ventilation and air-conditioning (HVAC) system during the summer when only cooling is necessary; this work naturally extends to heating. In this section we present the high-level HVAC model and control problem formulation. Details for setting up experiments are presented in Section \ref{results}.

\subsection{Building Thermal Dynamics}\label{mdl-building}

Studying building HVAC control requires a building thermal dynamics model for 1) applying model-based control methods, 2) generating data to train learning-based control methods, and 3) evaluating any control method. We choose the discrete-time model for a $z$-zone building
\begin{equation}\label{eqn:bilinear-model}
    \Tzones(t+1) = \mathbf{A}\Tzones(t) + \mathbf{B} \operatorname{diag}(\mathbf{\dot{m}}(t)) (\mathbf{1}_z \Tsupply(t) - \Tzones(t)) + \mathbf{G} \w(t),
\end{equation}
where $t$ is the control step index, $\mathbf{T}$ is the indoor temperature in each zone $i\in\{1,...,z\}$: $\mathbf{T} = [T^1, ..., T^z]^\top$, $\mathbf{\dot{m}}$ is the mass flow rate of air to each zone: $\mathbf{\dot{m}} = [\dot{m}^1, ..., \dot{m}^z]^\top$, and $\Tsupply$ is the temperature of the air supplied to all zones. The control action $\u \in \mathbb{R}^{z+1}$ consists of $\mathbf{\dot{m}}$ and $\Tsupply$. The exogenous input $\w \in \mathbb{R}^{z+1}$ consists of the outdoor temperature and per-zone solar heat gain. The time-invariant system matrices $\mathbf{A}\in\mathbb{R}^{z\times z}$, $\mathbf{B}\in\mathbb{R}^{z\times z}$, and $\mathbf{G}\in\mathbb{R}^{z\times (z+1)}$ are identified from a higher fidelity EnergyPlus~\cite{energyplus} model, as done in \cite{chintala2015automated}. Frequently used terms are defined again in the nomenclature table for easy referencing.

The term $\operatorname{diag}(\mathbf{\dot{m}}(t)) \mathbf{1}_z \Tsupply$ is a product among elements of $\u$ so the building model is bilinear and thus nonlinear and non-convex\footnote{$\operatorname{diag}(\mathbf{\dot{m}}(t))$ denotes a diagonal matrix with the elements of $\mathbf{\dot{m}}(t)$ along the diagonal and $\mathbf{1}_z$ denotes a $z\times1$ vector of ones.}. To simplify notation, we subsequently refer to the dynamics as $\mathcal{F}$, where
\begin{equation}\label{eqn:generic-model}
     \x(t+1) = \mathcal{F}\left( \x(t), \u(t), \w(t) \right),
\end{equation}
and $\x(t)$ is the building state, obtained via a linear transformation of $\Tzones$.

\subsection{Power Consumption and Problem Cost}\label{modeling:cost}

The HVAC power consumption is the sum of chiller and fan power,
\begin{align}
    \label{eq:chiller-power}
    \ptotal(t) &= \pchill(t) + \pflow(t), \\
    \pchill(t) &= \frac{\mathbf{1}_z^{\top} \mathbf{\dot{m}}(t)}{\COP}\left(\Tout(t) - \Tsupply(t)\right), \\
    \pflow(t) &= k_1\left(\mathbf{1}_z^{\top} \mathbf{\dot{m}}(t) \right)^3 + k_2\:,
\end{align}
where $\COP$ is the chiller's coefficient of performance, $\Tout$ is the outdoor temperature, and $k_1$ and $k_2$ are known fan parameters.

The cost $c$ of one control step $t$ is the weighted sum of the electricity cost and a penalty for violating the thermal comfort of building residents,
\begin{equation}\label{eqn:cost-model}
    c(\x(t), \u(t), \w(t)) = \lambda(t) \ptotal (t) \tau + \mu \sum_{i=1}^z \underset{[\underline{T}(t), \overline{T}(t)]}{\mathcal{P}}\left(T^i(t)\right),
\end{equation}
where $\lambda$ is the price of electricity discussed in section \ref{modeling:DR} and $\tau$ is the duration of a control step. The penalty uses the band deviation function which, for a range of desired values $\left[ \underline{x}, \overline{x} \right]$, is
\begin{equation}  \label{eq:band-deviation-penalty}
    \underset{[\underline{x}, \overline{x}]}{\mathcal{P}}\left(x\right) := \left[\underbrace{\max\left(0, \underline{x} - x\right)}_{\text{Lower constraint violation}} + \underbrace{\max\left(0, x - \overline{x}\right)}_{\text{Upper constraint violation}}
    \right]^2.
\end{equation}
The thermal comfort band for our control problem, $\left[ \underline{T}(t), \overline{T}(t) \right]$, varies in time. Deviations from the band are penalized at a fixed per-unit price $\mu$.

\subsection{Demand Response (DR) Programs}\label{modeling:DR}

To thoroughly compare controller performance for building control under DR, we consider three common DR programs: time-of-use pricing (TOU), real-time pricing (RTP), and power-constrained (PC). Each program results in a different price for electricity $\lambda(t)$. TOU uses a preset price for each time of day, RTP uses a dynamically-varying price that depends on the hour-by-hour wholesale electricity market, and PC uses a constant price but includes an extra term in ~\eqref{eqn:cost-model} to penalize consumption above a given power limit during a power-constraint event. Details of each DR program are presented in Section \ref{subsubsec-scenario-setup}.

\subsection{Building Control Problem Formulation}  \label{subsec-problem-formulation}

Building control minimizes the cost to keep the zone temperatures within a comfort band over a $N$-step horizon while satisfying the building thermal dynamics $\mathcal{F}$ and constraints on HVAC actuation:
\begin{subequations}\label{eqn:ocp}
\begin{align}
    \min_{\u_0,...,\u_{N-1}} \quad C = &\sum_{t=0}^{N-1} c(\x_t, \u_t, \w_t) \label{eqn:ocp-cost}\\
    \text{s.t.}\quad & \x_{t+1} = \mathcal{F}(\x_t, \u_t, \w_t), \label{eqn:dynamics}\\
    & \x_t \in \mathcal{X}, \u_t \in \mathcal{U},\\
    & \x_0 = \x(0). \label{eqn:dynamics-x0}
\end{align}
\end{subequations}
The initial building state is $\x(0)$. In general, we constrain $\x$ and $\u$ to lie in sets $\mathcal{X}$ and $\mathcal{U}$, respectively. In this work, $\x$ is unconstrained, $\mathcal{X}=\mathbb{R}^{z}$, and each element in $\u$ is constrained by upper and lower bounds, making $\mathcal{U}$ a box polyhedron in $\mathbb{R}^{z+1}$. 

As the first step for comparing a spectrum of building controllers and to focus on their core capabilities, we assume:

\begin{itemize}
    \item[\textbf{A.1}] There is no error in the building model. This translates to using the same model in our controllers and building simulation. 
    \item[\textbf{A.2}] Forecasts are accurate up to a $K$-step lookahead. This means at $t$ the disturbances $\{\w(t), \w(t+1), ..., \w(t+K-1)\}$ are accurately known. We investigate how the choice of $K$ affects each controller.
\end{itemize}

Given our baseline comparison, future work can investigate the effect of model and forecast error, such as by integrating \cite{lazos_forecasting_2014}. By first comparing the control methods without these errors, we make it possible for future work to isolate to what extent one method outperforms another due to differences in baseline performance, sensitivity to model error, and sensitivity to forecast error.

\subsection{Measured signals}  \label{subsec-feature-vector}

All controllers in this study rely on signals from the building and environment. To enable a fair comparison, at step $t$, all controllers use the following signals for decision-making:
\begin{itemize}
    \item current zone temperature $\Tzones(t)$ or system state $\x(t)$,
    \item current time step $t$, 
    \item $K$-step forecasts of disturbances from the current time, $\w_K(t)=\{\w(t), \w(t+1), ..., \w(t+K-1)\}$,
    \item DR-specific information $\Lambda$, as detailed in Section \ref{results}.
\end{itemize}
As will be presented in Section \ref{control}, some control methods use these signals directly in an optimization formulation, while others transform them into a feature vector, $\s(t) = g(\x(t), \w_K(t), \Lambda(t), t)$. Details of the mapping function $g(\cdot)$ are given in Section \ref{subsubsec-learning-setup}.

\section{Control Methods}\label{control}

\begin{figure*}[]
\centering
\begin{framed}
  \includegraphics[width=0.95\linewidth]{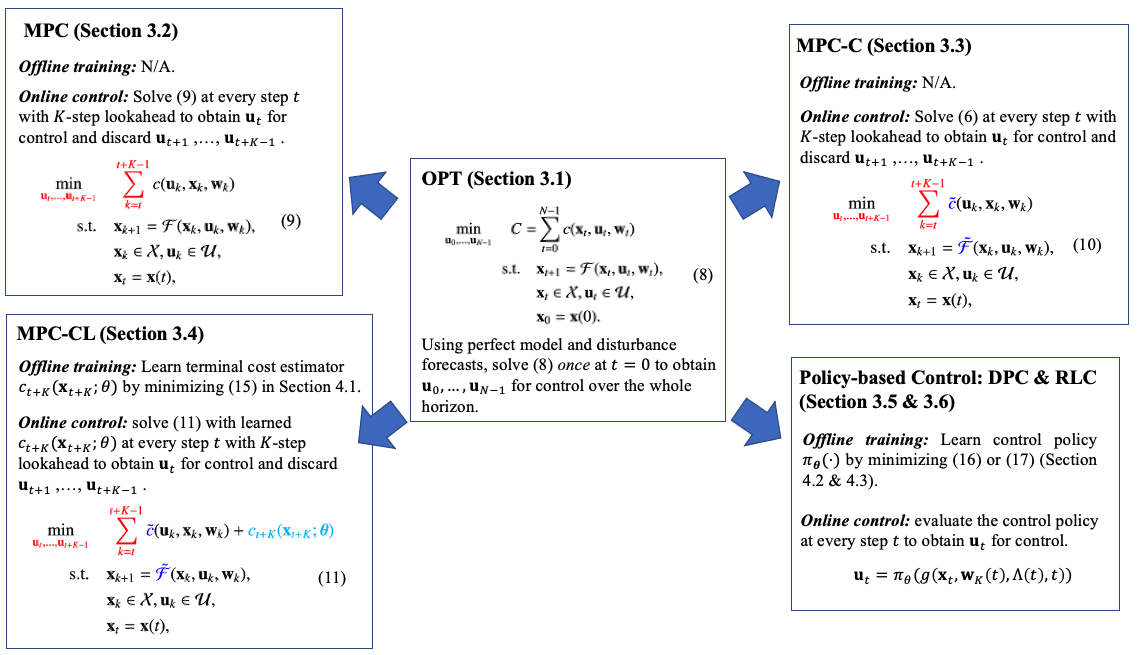}
\end{framed}
\caption{An overview of controllers compared in this study. All controllers aim to solve \eqref{eqn:ocp}, but instead solve an approximate version of this problem. The modifications made in these approximations are highlighted with different colors.}
\label{fig-controller-relation}
\end{figure*}

\subsection{Open-loop Optimal Control Baseline (OPT)}\label{control:mpco}

The control problem~\eqref{eqn:ocp} can be solved in an open-loop manner with no feedback by solving the problem once at time step $t=0$ and then applying the resulting control actions to the system --- we do so using the numerical solver IPOPT \cite{wachter2006}. Since we assume there is no error in the building model and forecasts of exogenous inputs (recall assumptions \textbf{A.1} and \textbf{A.2}), this approach gives the optimal control actions. As such, we refer to it as the optimal control baseline (OPT) and use it to evaluate other control methods, which are approximations to OPT. 

On a real system with modeling and forecast errors, solving the open-loop problem~\eqref{eqn:ocp} would work poorly, due to accumulation of errors. The subsequent sections present the controllers studied in this paper, which approximate OPT as shown in Figure \ref{fig-controller-relation}. The approximation methods vary, in particular in their use of the building model and learning.

\subsection{Model Predictive Control (MPC)}\label{control:mpc}

Where OPT solves the control problem~\eqref{eqn:ocp} once over the full control horizon, model predictive control (MPC) instead solves an analogous problem at every control step index $t$ over a lookahead horizon $K \leq N$. This updates the optimal control action online to account for any modeling mismatches and respond to updated disturbance forecasts. The MPC problem is
\begin{subequations} \label{eqn:nmpc}
\begin{align}
    \min_{\u_t,...,\u_{t+K-1}} \quad & \sum_{k=t}^{t+K-1} \color{black} c(\x_k, \u_k, \w_k) & \\
    \text{s.t.}\quad & \x_{k+1} = \mathcal{F}(\x_k, \u_k, \w_k),\\
    & \x_k \in \mathcal{X}, \u_k \in \mathcal{U},\\
    &\x_t = \x(t),
\end{align}
\end{subequations}
where $k$ is the lookahead step index and $t$ is the control step index. MPC solves problem~\eqref{eqn:nmpc} once for each value of $t$ from $0$ to $N-1$. 
The non-convex optimization problem \eqref{eqn:nmpc} is solved online using IPOPT; once the solution, $\{\u^*_t,...,\u^*_{t+K-1}\}$, is found, the first action is applied to the system ($\u(t) = \u^*_t$.) Then the new state, $\x(t+1)$, is observed and problem~\eqref{eqn:nmpc} is solved again, starting from time $t+1$. 

MPC does not give the same control actions as OPT, because the objective functions in problems~\eqref{eqn:ocp} and ~\eqref{eqn:nmpc} are not the same; MPC is an approximation to OPT, meaning it gives suboptimal control performance. Choosing the total number of lookahead steps $K$ requires balancing control performance and computational cost. When we neglect error, a larger $K$ improves the approximation to OPT and thus improves control performance but also increases computational cost. Note that when error is included, increasing $K$ does not necessarily improve control performance since it improves the approximation to OPT but increases accumulation of error. 

Beyond the choice of $K$, some MPC implementations aim to improve control performance by adding a terminal cost function, $c_{t+K}(\x_{t+K})$, which models the cost of ending in state $\x_{t+K}$. The optimal terminal cost can in theory be found through dynamic programming ~\cite{bertsekas_4thed_2012}, where a sequence of one-step control problems are solved backwards in time (starting with $t=N-1$) using the Bellman equation. However, even for discrete-time finite horizon problems such as \eqref{eqn:ocp}, DP requires optimizing over the space of functions and is thus an infinite-dimensional problem. This makes it challenging to determine an appropriate $c_{t+K}(\x_{t+K})$; for simplicity, we do not include a terminal cost in our MPC implementation. However, we explore using learning to find the terminal cost in Section \ref{control:mpccl}.

\subsection{Convex Model Predictive Control (MPC-C)}\label{control:mpcc}

Besides choosing a small lookahead horizon $K$, one can reduce the computational cost of MPC by using a simpler model, such as a convex one. We denote MPC with a convex building model as MPC-C. The convex model is the same as that described in Section \ref{mdl-building} except that the non-convex bilinear terms in $c$ and $\mathcal{F}$ are linearized around the system’s current operating point. The MPC-C problem is 
\begin{subequations} \label{eqn:mpc-c}
\begin{align}
    \min_{\u_t,...,\u_{t+K-1}} \quad & \sum_{k=t}^{t+K-1} \color{black} \tilde{c}(\x_k, \u_k, \w_k) & \\
    \text{s.t.}\quad & \x_{k+1} = \tilde{\mathcal{F}}(\x_k, \u_k, \w_k),\\
    & \x_k \in \mathcal{X}, \u_k \in \mathcal{U},\\
    &\x_t = \x(t),
\end{align}
\end{subequations}
where $\tilde{c}$ and $\tilde{\mathcal{F}}$ are the convex cost and dynamics.

\subsection{Convex Model Predictive Control with Learned Terminal Cost (MPC-CL)}\label{control:mpccl}

Reinforced MPC (RL-MPC) is a hybrid control approach that addresses the challenge of determining an appropriate terminal cost function, $c_{t+K}(\x_{t+K})$; this challenge was described in Section \ref{control:mpc}. RL-MPC includes the terminal cost in MPC by parameterizing the terminal cost, $c_{t+K}(\x_{t+K}; \param)$, and learning the optimal parameters $\param^*$ offline through training. During training, for a batch of data and control timestep one solves the MPC problem over the lookahead horizon, evaluates the actual cost of taking the actions from the MPC solution, finds the gradient of this cost with respect to the parameters by differentiating through the MPC problem, and then takes a negative gradient step to update $\param$ in the direction that decreases cost. Online, RL-MPC solves the MPC problem with the learned terminal cost. 

Differentiating through the MPC problem (and thus also through the building model) is a non-trivial task. To do so, we use the CVXPYLAYERS open source library \cite{agrawal2019differentiable}. This library requires that our optimization problem --- and thus our building model --- be convex. We therefore use the same convex dynamics $\tilde{\mathcal{F}}$ and cost function $\tilde{c}$ as in MPC-C to obtain our control problem
\begin{subequations}  \label{eqn:lmpc}
\begin{align}
    \min_{\u_t,...,\u_{t+K-1}} \quad & \sum_{k=t}^{t+K-1} \tilde{c} \color{black}(\x_k, \u_k, \w_k) + c_{t+K}(\x_{t+K}; \param) \label{eqn:lmpc-cost}\\
    \text{s.t.}\quad & \x_{k+1} = \tilde{\mathcal{F}}\color{black}(\x_k, \u_k, \w_k),\\
    & \x_k \in \mathcal{X}, \u_k \in \mathcal{U},\\
    &\x_t = \x(t).
\end{align}
\end{subequations}

We refer to this approach as convex MPC with learned terminal cost (MPC-CL) to distinguish our formulation of RL-MPC from previous works, such as \cite{kowli2012coordinating} and \cite{arroyo2022reinforced}. We build upon these works by using a more general form for the terminal cost than \cite{kowli2012coordinating} and a less computationally expensive training method than \cite{arroyo2022reinforced}. We describe our training method to learn the parameters $\param$ that accurately estimate the terminal cost in Section~\ref{training:cpl}. 

\subsection{Reinforcement Learning Control (\rl)}\label{control:rl}

In contrast to MPC approaches, model-free RL does not optimize the control actions $\u_t$ online directly with knowledge of the building model. Rather, the building model is a black box to which inputs (control actions $\u_{t}$) can be given and from which measurements (states $\x_{t}$) can be taken offline to learn a parameterized policy $\pi_{\param}(\cdot)$ for generating control actions. Note that the modifier ``model-free'' means that no model is used during online control or explicitly used (e.g. differentiated through) during training. In engineering applications such as building control, a black box model is used during training, since training on a real system could lead to expensive or unsafe actions. 

During online control, RL simply evaluates the learned policy on an observation $\s(t)$ to generate the control action:
\begin{equation}\label{eqn:RL-policy}
    \u(t) = \pi_{\param} (\s(t))\:,
\end{equation}
where the feature vector $\s(t)$ was introduced in Section \ref{subsec-feature-vector} and the policy parameters $\param$ are learned offline during training\footnote{We use the same notation $\param$ to represent a learnable parameter vector in Sections \ref{control:mpccl}-\ref{control:dpc}, respectively; the dimension of $\param$ depends on the section.}.

Training aims to find the policy that optimizes \eqref{eqn:ocp}. However, it only searches over polices with the parametrization $\param$, whereas the optimal solution, OPT, effectively searches over all polices. This means RL, like all methods in this study, is an approximation of OPT.

Many forms of policies and RL algorithms exist. We use a neural network (NN) as our policy; therefore, $\param$ collects the weights and biases of the policy network. 
The NN architecture is described in Section \ref{ssec:training}. We use an actor-critic algorithm, proximal policy optimization (PPO) \cite{schulman2017proximal}. The details of this method become relevant during training, as explained in Section~\ref{training:rl}. 

\subsection{Differentiable Predictive Control (DPC)}\label{control:dpc}

The ability of RL to use a black box model with little constraint on the model form makes it very flexible. However, RL is not sample-efficient, meaning it requires many input-output samples for the policy parameters to converge. This inefficiency comes from estimating the gradient of the cost with respect to the parameters based on samples. When the model is directly accessible and differentiable (i.e. it is a white box model), one can write the multi-step control cost, \eqref{eqn:ocp-cost}, as a function of the policy parameters $\param$, forming a computational graph. Then the policy gradient can be computed by differentiating through the computational graph using deep learning software frameworks. This approach is called differentiable predictive control (DPC) ~\cite{drgona_DPC_2021}\footnote{While the term ``predictive control'' suggests an explicit predictive model is used online, as in MPC, predictions are actually implicitly embedded in the policy---no explicit model is used online.}. As with RL, during online control, DPC simply evaluates the policy $\pi$ to generate the control action, as expressed in \eqref{eqn:RL-policy}.

For a fair comparison, the DPC policy uses the same NN architecture as RLC, detailed in Section \ref{ssec:training}. Policy parameters $\param$ are trained offline by differentiating through the computational graph, as detailed in Section~\ref{training:dpc}. The comments regarding the suboptimality of RLC also stand for DPC.

\subsection{Rule-Based Control (RBC)}\label{control:rbc}

Finally, we include the industry-standard approach for building control under DR programs: rule-based control (RBC). RBC is a heuristic approach developed by a building expert. Our RBC approximates the optimal control problem~\eqref{eqn:ocp} with the following strategy: ``Prior to high-cost events, pre-cool the building while maintaining thermal comfort. Otherwise, follow a temperature set point with minimal control effort." This results in a set of cooling rules based on the time of day, the building system $\mathcal{F}$, the constraints $\mathcal{X}$ and $\mathcal{U}$, the expected energy pricing signal, and the expected impact of disturbances. These rules were found by conducting a grid search over a range of controller configurations and selecting those which give the lowest cost while respecting the constraints. We refer the reader to our code \cite{lbc_code} for the implementation details. 

\section{Training Learning-Based Controllers}\label{training}

Of the methods considered, MPC-CL (Section~\ref{control:mpccl}), RL (Section~\ref{control:rl}), and DPC (Section~\ref{control:dpc}) require parameter optimization, called training. During training, each method defines a loss function $\mathcal{L}(\param)$ and uses it to perform gradient-based updates of the parameters. For example, the stochastic gradient descent (SGD) algorithm uses the update 
\begin{align} \label{eqn:parameter-learning}
    \param_{i+1} = \param_{i} - \alpha \nabla_{\param_i} \mathcal{L}(\param_i),
\end{align}
where $\alpha$ is the learning rate that controls the size of the update, $\theta_i$ is the current parameter estimate, and $\theta_{i+1}$ is the new parameter estimate. While SGD is useful for conceptual understanding, in practice we use Adam optimization \cite{kingma2014adam} as it typically converges more stably and quickly than SGD. Next, we discuss how each of our learning-based control methods defines $\mathcal{L}(\param)$.

\subsection{\cvxpylayer}\label{training:cpl}

The MPC-CL controller needs to be trained to obtain $\param$ for the terminal cost estimator in \eqref{eqn:lmpc-cost}. In this study, $c_{t+K}(\x_{t+K}; \param)$ is chosen to be linear in the system state. To account for the time-dependence of DR programs without explicitly including the full forecast of exogenous inputs, different parameters vector $\param_j$ for each time interval are used so that the terminal cost is time dependent: $c_{t+K}(\x_{t+K};\param_j) := \param_j^T \x_{t+K}$. The time interval $j$ can be different from $k$; for example, $k$ could index five-minute intervals while $j$ indexes one-hour intervals. Other parametrizations of the terminal cost are also possible, such as a quadratic, $c_{t+K}(\x_{t+K}; \bm{M}_j, \param_j)=\x_{t+K}^T \bm{M}_j \x_{t+K} + \param_j^T \x_{t+K}$, so long as the form is convex. We use a linear form, because we found it provided sufficient information to allow the MPC-CL controller to plan for DR events beyond the lookahead horizon. In particular, one of our main contributions is identifying that this simple method can lead a learnable, convex MPC model (MPC-CL) to significantly outperform exact, non-convex MPC (MPC) for the same lookahead horizon, as shown in Section \ref{results}.

The training process for the parameters $\param_j$ consists of two steps at each training iteration $i$. First, the MPC problem (\ref{eqn:lmpc}) is solved at each time $t=0,\dots,N-1$ using current values for $\param_{j,i}$, and the optimized first action $\mathbf{u}^{(b)}_t|_{\param_{j,i}}$ is applied to the simulation. This is done for $|\mathcal{B}|$ randomly drawn scenarios, indexed by $b$, where each scenario is one day of operation and $\mathcal{B}$ is a batch of scenarios. The realized cost for each control step is accumulated into a loss and averaged over the batch:
\begin{align} \label{eqn:cpl-loss}
    \mathcal{L}(\param_{j,i}) = \frac{1}{|\mathcal{B}|}\sum_{b \in \mathcal{B}}  \sum_{t=0}^{N-1} c\left(\mathcal{F}(\mathbf{x}_{t-1}^{(b)}, \mathbf{u}_{t-1}^{(b)}|_{\param_{j,i}}, \mathbf{w}_{t-1}^{(b)}),\:\mathbf{u}^{(b)}_t|_{\param_{j,i}}, \mathbf{w}_{t}^{(b)}\right).
\end{align}
For the $t=0$ term, $\mathcal{F}(\mathbf{x}_{-1}^{(b)}, \mathbf{u}_{-1}^{(b)}|_{\param_{j,i}}, \mathbf{w}_{-1}^{(b)})$ is replaced with the initial condition $\x(0)$. Second, the gradient $\nabla_{\param_{j,i}} \mathcal{L}$ is used to compute updates to the controller parameters, $\param_{j, i}\xrightarrow[]{\text{update}}\param_{j, i+1}$, using the Adam update~\cite{kingma2014adam} (similar to \eqref{eqn:parameter-learning}). This procedure is repeated by drawing batches from training scenarios without replacement. Once the days have been exhausted (the epoch is complete), all days reenter the sampling pool and the next epoch begins using the final parameters from the previous epoch. Note that the learned $c_{t+K}(\x_{t+K};\param_j)$ only indirectly influences the loss function~\eqref{eqn:cpl-loss} through its effect on the computed control action at every time step.

Because the scenarios $b$ within a batch $\mathcal{B}$ do not affect each other, the loss function~\eqref{eqn:cpl-loss} (and relevant gradients) can be evaluated in parallel before computing the parameter update. As such, we organize training to learn the parameters using PyTorch batching. However, the requirement to solve a convex optimization problem at each forward pass prohibits GPU acceleration because the underlying solvers are not set up to use GPUs.

\subsection{\rl}\label{training:rl}

The RL algorithm PPO is on-policy, meaning that the current policy, $\pi_{\param_i} (\cdot)$, is used to generate data that is subsequently used to update it. Given the current RL observation $\s_t$, an action $\u_t$ is drawn from $\pi_{\param_i} (\cdot)$ and passed through the model to produce both a subsequent state $\mathbf{x}_{t+1}$ and a reward $r_t = -c(\x_t, \u_t, \w_t)$. A collection of observation-action-reward-next observation tuples is called the experience: $ \mathcal{E} = \{(\s, \mathbf{u}, r, \s')^i\}_{i\in[0, ..., |\mathcal{E}|]}$. Using the experience collected in each training iteration, PPO defines the loss function as
\begin{equation}
\label{eqn:rl-loss}
\begin{split}
    \mathcal{L} (\param_{i}) = \frac{1}{|\mathcal{E}|} \sum_\mathcal{E} \left\{ \min \left(\eta_t(\param_{i})\hat{A}_t(\s_t, \u_t), \right.\right. \\
    \left.\left.\operatorname{clip}(\eta_t(\param_{i}), 1-\beta, 1+\beta)\hat{A}_t(\s_t, \u_t)\right)\right\}.
\end{split}
\end{equation}
Here, $\eta_t(\param_{i})$ is the ratio of the probability of the current policy with parameters $\param_{i}$ outputting action $\mathbf{u}_t$ to the probability of the old policy with parameters $\param_{i-1}$ outputting $\mathbf{u}_t$ when observing $\s_t$: 
$$
    \eta_t(\param_{i})=\frac{\pi_t(\u_t \,|\, \s_t; \param_{i})}{\pi_t(\u_t \,|\, \s_t; \param_{i-1})}\ .
$$
$\beta$ is a hyperparameter for clipping. The number of sets included in the experience, $|\mathcal{E}|$, is also a hyperparameter. The $\operatorname{clip}(\cdot)$ and $\min(\cdot)$ operators penalize large policy updates, encouraging the policy parameters to remain in a trust region, as explained in \cite{schulman2017proximal}. The advantage estimate $\hat{A}_t(\s_t, \u_t)$ approximates the difference between the reward from taking action $\u_t$ when observing $\s_t$ and the expected value of observing $\s_t$ and then taking an action according to the current policy. We use the RLLib default advantage estimator, the Generalized Advantage Estimator (GAE) from \cite{schulman2018}. GAE estimates the expected value using a NN called the critic NN that is separate from the policy NN, called the actor NN. The full PPO loss function includes additional terms, see (9) in \cite{schulman2017proximal}, which we omit in \eqref{eqn:rl-loss} for conciseness. Note that, unlike the MPC-CL loss~\eqref{eqn:cpl-loss} and DPC loss~\eqref{eqn:dpc-loss} defined in the next section, the RL loss~\eqref{eqn:rl-loss} does not directly use the system dynamics, meaning $\mathcal{F}$ does not appear in the loss function. This is in line with the model-free nature of the RL algorithm. Our implementation of RL clips the control action to satisfy the constraints; future work should include a penalty for constraint violations.

During learning, the loss~\eqref{eqn:rl-loss} is differentiated with respect to the parameter $\param_{i}$, yielding $\nabla_{\param_{i}} \mathcal{L} (\param_{i})$ to update $\param$. 

RL training can be effectively parallelized; how to best do so depends strongly on the problem and resources. Generally speaking, RL training consists of two interdependent process: sampling---experience collection using current policy---and learning---policy parameter updates via backpropagation. In this study, we leverage the scalable RL framework RLLib \cite{liang2018rllib} to train the control policy using multiple CPUs for sampling and learning.

\subsection{\dpc}\label{training:dpc}

Although the structure of DPC is similar to that of RLC, the availability of a differentiable model makes training, in particular defining the loss, simpler. As in MPC-CL, we take $\mathcal{B}$ randomly drawn samples, apply an action, and use our model to calculate the realized cost; unlike MPC-CL, here the action comes from the DPC policy. As in MPC-CL the realized cost for each control step is accumulated into a training loss and averaged over the batch:
\begin{equation}
\label{eqn:dpc-loss}
\begin{split}
    \mathcal{L}(\param_{i}) = \frac{1}{|\mathcal{B}|}\sum_{b \in \mathcal{B}}  \sum_{t=0}^{N-1} \left\{ c\left(\mathcal{F}(\mathbf{x}_{t-1}^{(b)}, \mathbf{u}_{t-1}^{(b)}|_{\param_{i}}, \mathbf{w}_{t-1}^{(b)}),\:\mathbf{u}^{(b)}_t|_{\param_{i}}, \w_t^{(b)}\right) \right. \\ 
    \left. + \rho h\left(\mathbf{u}^{(b)}_t|_{\param_{i}}\right) \right\}
\end{split}
\end{equation}
where $\mathbf{u}^{(b)}_t|_{\param_{i}}=\pi_{\param_{i}}(g(\x^{(b)}_t, \w^{(b)}_{t, K}, \Lambda^{(b)}_t, t))$ is the action determined by the control policy and $\rho$ is a non-negative, tunable hyperparameter.

Unlike MPC-CL, DPC does not inherently account for constraints. Instead, the DPC loss function is augmented with a penalty for constraint violations, $h(\u) = \sum_{i=1}^{z+1} \underset{[\underline{u}, \overline{u}]}{\mathcal{P}} \left( u^{i} \right)$, where $\mathcal{P}$ is defined in \eqref{eq:band-deviation-penalty}. To ensure the loss function is automatically differentiable, we use the $\tt{relu}$ activation function rather than the $\max$ operator in \eqref{eq:band-deviation-penalty}, as in \cite{drgona2020learning}. Similar to RLC training, the gradient $\nabla_{\param_i} \mathcal{L}$ is used to update $\param$.

Because DPC training is essentially differentiating through the computational graph developed by unrolling the control horizon, it readily lends itself to GPU acceleration. Similarly to MPC-CL, each batch $\mathcal{B}$ consists of a number of scenarios of simulated data that can be evaluated in parallel in both forward and backward (gradient) passes.

\section{Experiments and Results}\label{results}

\subsection{Experiment Setup}

The above-mentioned controllers are applied to control the HVAC system of a building under DR programs. Section \ref{subsubsec-scenario-setup} details the experiment setup, including building model parameters, data, and DR programs. The training process is described in Section \ref{subsubsec-learning-setup} and the software implementation in Section \ref{subsubsec-software-implementation}.

\subsubsection{Experiment Setup} \label{subsubsec-scenario-setup}

\begin{figure}[]
\centering
\includegraphics[width=0.5\linewidth]{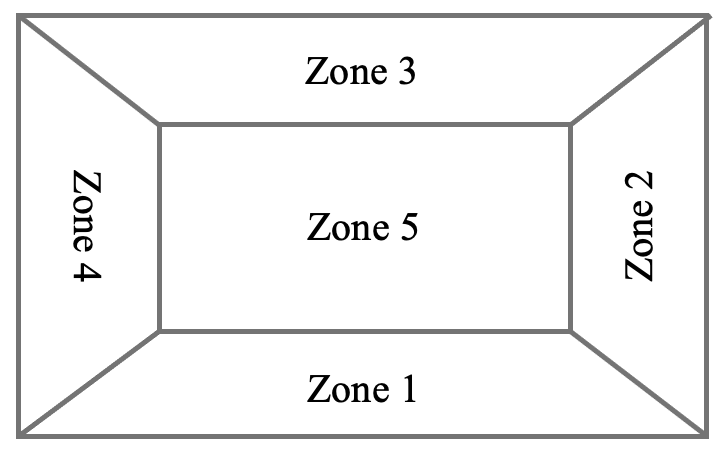}
\caption{Floor plan for the five-zone small office building.}
\label{fig-five-zone-building}
\end{figure}

We use a reduced order model (ROM) derived from the U.S. Department of Energy (DOE) reference small office building \cite{buildingmodel}. This building consists of five thermal zones ($z=5$), as shown in Figure \ref{fig-five-zone-building}. Building thermal dynamics are given in \eqref{eqn:bilinear-model} and the associated parameters are specified in Table \ref{table-office-building-parameters}. The controllers aim to minimize the daily operation cost as defined in \eqref{eqn:ocp-cost} for control steps of duration $\tau=5$ minutes over all $N=288$ control steps in a day.

\begin{small}
\begin{table}[]
\centering
\footnotesize
\caption{Parameters for the five-zone building}
\label{table-office-building-parameters}
\begin{tabular}{c|l}
\specialrule{.15em}{.075em}{.075em}
\multirow{3}{*}{\begin{tabular}[c]{@{}c@{}}\textbf{HVAC Operating}\\ \textbf{Constraints}\end{tabular}} 
& $\dot m_t^i \in [0.22, 2.2], i \in \{1,...,4\}$ (unit: $kg/s$), \\
& $\dot m_t^5 \in [0.32, 3.2]$ (unit: $kg/s$) \\
& $\Tsupply \in [10.0, 16.0]$ (unit: $^\circ C$) \\ \hline
\textbf{HVAC Power Parameters}
& $\text{COP}= 3.0$, $k_1=0.0076$, $k_2=4.8865$ \\ \hline
\begin{tabular}[c]{@{}c@{}}\textbf{Indoor Comfort}\\ 
\textbf{Temperature Band}\end{tabular}
& \begin{tabular}[c]{@{}l@{}} $[\underline{T}^i, \overline{T}^i] = [20.0, 24.5]$ (7:00 - 18:00)\\ 
$[\underline{T}^i, \overline{T}^i] = [16.0, 28.0]$ (Other time)\end{tabular} \\ \specialrule{.15em}{.075em}{.075em}
\end{tabular}
\end{table}
\end{small}

To simulate building operation, we use exogenous data from July and August, including outdoor temperature and solar heat gain for each zone. Data from July are used to train learning-based controllers; data from August are used to test all controllers.

To comprehensively compare these building controllers, we test them on three common DR programs, two price-based and one incentive-based:
\begin{enumerate}
    \item Time-of-use (TOU): Electricity price varies by time of day according to a predetermined schedule. The price is either peak or off-peak, where the peak price is 2-10x higher than the off-peak price, as seen in both real-life \cite{sdge2022tou} and research \cite{brandi2022comparison}. In this paper, the following price schedule is used:
    \begin{equation}
        \lambda(t)=\left\{\begin{array}{ll} {10} & {(\text {peak: 12:00 - 18:00})} \\ {1} & {\text{(off-peak: 18:00 - 12:00 next day)}}
    \end{array}\right.
    \end{equation}
    \item Real-time price (RTP): Electricity price changes every hour based on the RTP from the wholesale market, which is known by the end of the hour. The day-ahead price (DAP) is given at the end of the previous day and provides a prediction of the next day's RTP. We use the RTP and DAP data from a real-world RTP program \cite{comed2022rtp}.
    \item Power constrained (PC): Customers receive a lower electricity price for participating in the program. In exchange, during specified times called DR events, their building must reduce its power below a predetermined limit for a period of time---otherwise a penalty is applied. The limit is typically calculated from the building's baseline consumption and load reduction potential. In this study, we use the limit
    \begin{equation}  \label{eq-pc-power-limit}
        P^{\text{limit}}(t)=\left\{\begin{array}{ll} {15 \text{ kW}} & {(\text {DR event: 12:30 - 16:30})} \\ {25 \text{ kW}} & {\text{(other times)}}
    \end{array}\right.
    \end{equation}
    The cost defined by \eqref{eqn:cost-model} is augmented by the penalty $\nu \underset{[0, P^{\text{limit}}(t)]}{\mathcal{P}}\left(P^{\text{total}}(t)\right)$, where $\mathcal{P}$ is defined in \eqref{eq:band-deviation-penalty} and the scaling factor is $\nu=10$. The electricity price is constant, $\lambda(t) = 1$.
\end{enumerate}

\subsubsection{Learning Setup}\label{subsubsec-learning-setup}

The RLC and DPC policies take in an observation vector, $\s(t) = g(\x(t), \w_K(t), \Lambda(t), t)$, as introduced in Section \ref{subsec-feature-vector}. The observation vector differs by DR program:
\begin{subequations}  \label{eqn:policy_obs}
\begin{align}
    &\s^{TOU}(t) = [\x(t), \Tzones(t), \mathbf{T}_K^{\text{out}}(t), \mathbf{G}_K^{\text{solar}}(t), \mathbf{em}(t)], \\
    &\s^{RTP}(t) = [\x(t), \Tzones(t), \mathbf{T}_K^{\text{out}}(t), \mathbf{G}_K^{\text{solar}}(t), \mathbf{em}(t), \bm{\lambda}^{\text{DA}}_K(t)], \\
    &\s^{PC}(t) = [\x(t), \Tzones(t), \mathbf{T}_K^{\text{out}}(t), \mathbf{G}_K^{\text{solar}}(t), \mathbf{em}(t), \mathbf{P}^{\text{limit}}_K(t)],
\end{align}
\end{subequations}
where $\mathbf{T}_K^{\text{out}}(t)$ and $\mathbf{G}_K^{\text{solar}}(t)$ are the $K$-step lookahead outdoor temperature and solar heat gain vectors. $\mathbf{em}(t)$ represents time embedding \cite[Eq.(6)]{zhang2019iot}. $\bm{\lambda}^{\text{DA}}_K(t)$ is the $K$-step lookahaead day-ahead electricity price signal and $\mathbf{P}^{\text{limit}}_K(t)$ is the $K$-step ahead power limit for PC programs.

During training, MPC-CL learns a terminal cost, and DPC and RLC each learn a control policy instantiated by a neural network (NN). To ensure a fair comparison between DPC and RLC, they share the same NN architecture. For TOU and RTP, the NN has hidden layers of size [512, 512]. For PC, the NN has hidden layers of a smaller size [32, 32] to avoid over-fitting. We conduct the training on the high-performance computing (HPC) system at the U.S. National Renewable Energy Laboratory (NREL), using compute nodes with the Dual Intel Xeon Gold Skylake 6154 (3.0 GHz, 18-core) processors. Training MPC-CL and DPC uses one compute node while training RLC is parallelized on two computing nodes.

\subsubsection{Software Implementation}\label{subsubsec-software-implementation}

\begin{figure}[]
\centering
\includegraphics[width=0.95\linewidth]{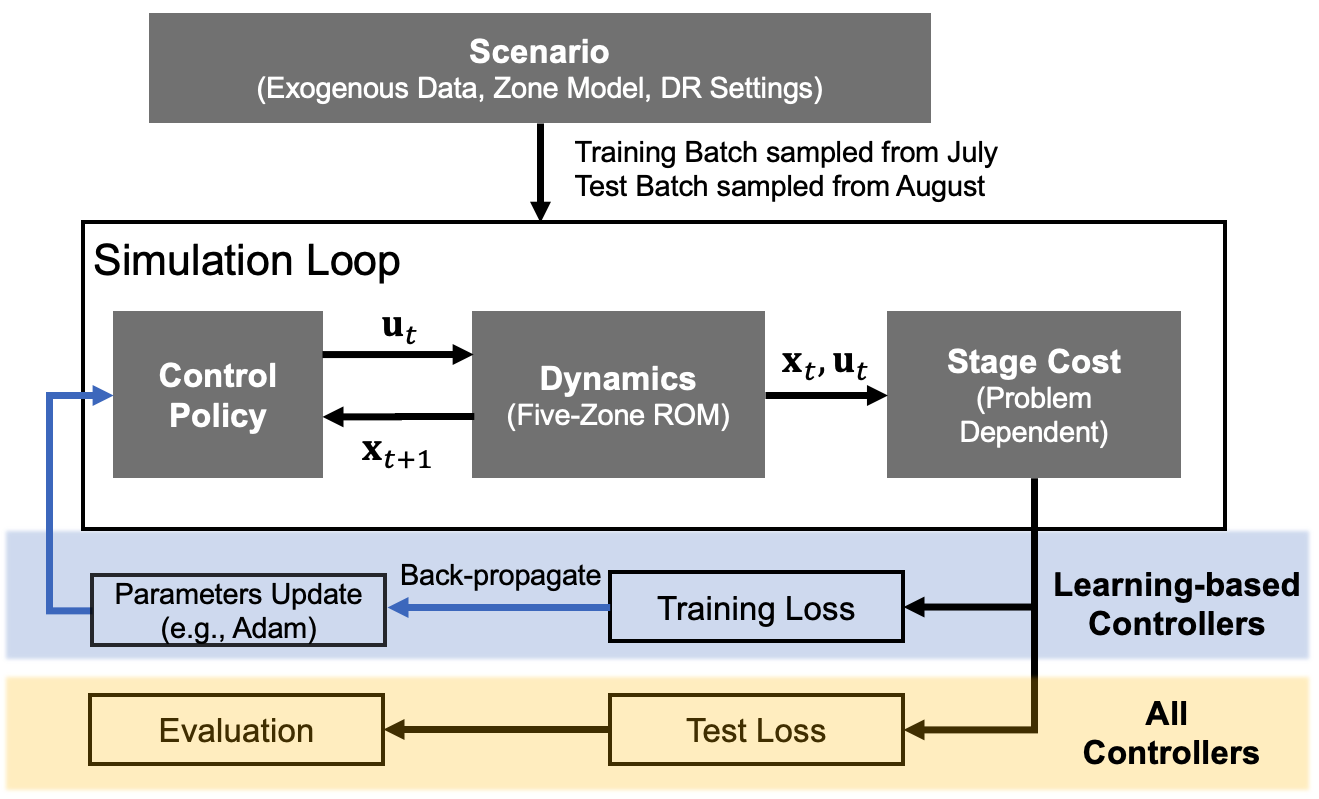}
\caption{The software pipeline used in this paper was designed to compare control methods with significantly different computational paradigms by abstracting out policy, dynamics, and cost modules.}
\label{fig-software-framework}
\end{figure}

To ensure a fair comparison of control methods and enable future comparisons to and extensions of our work, we develop a modular software framework, as shown in Figure \ref{fig-software-framework}. This design allows controllers to ``plug-and-play'' with the scenarios, dynamics, and training method, meaning any of these components can be changed without needing to change the controller implementation. Interested readers can use our software in their studies, including by customizing scenarios, system dynamics, and cost functions and comparing new control methods to those considered in this work. Our code is available open-source at \cite{lbc_code}. This code includes both non-convex and convex building models, which can be differentiated through; this feature is not available in existing building simulation and control suites. This feature of our code enables building control with hybrid methods, such as MPC-CL and DPC, which train by differentiating through a model.

\subsection{Training Performance for Learning-Based Controllers}\label{ssec:training}

\begin{figure}[]
\centering
\includegraphics[width=0.9\linewidth]{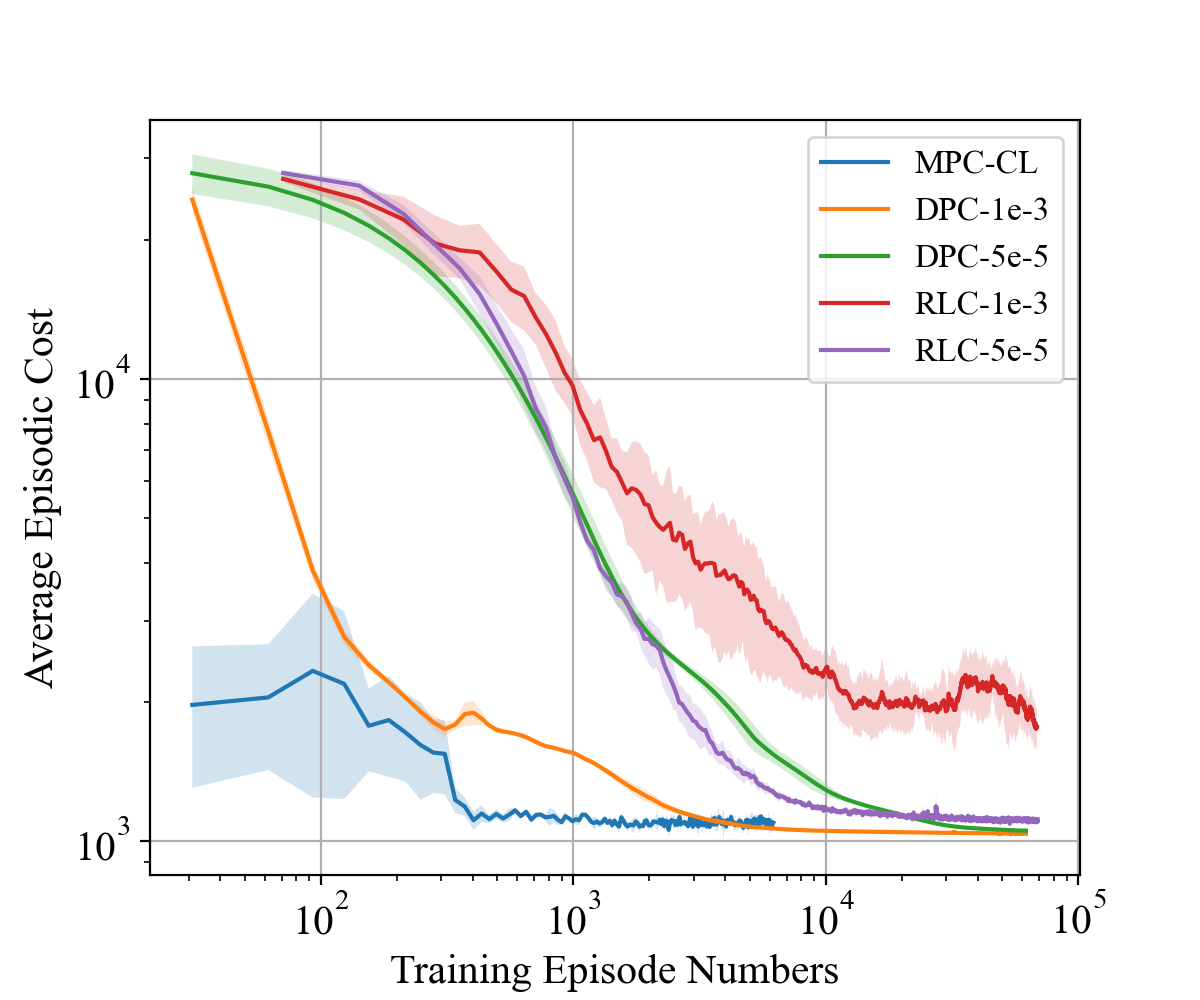}
\caption{Learning curves for training three learnable building control algorithms under a TOU program. All curves are clustered from experiments with three different lookahead horizons (15-min, 30-min, and 1-hr), where solid curves mark the mean and shaded areas the standard deviation. The ``1e-3'' and ``5e-5'' in the legend indicate learning rates used in DPC and RLC training, as described in Section \ref{ssec:training}. Both axes use log scaling. The learning curves are similar for RTP and PC programs.}
\label{fig-lc-comparison}
\end{figure}

\begin{small}
\begin{table*}
\centering
\footnotesize
\caption{Computational time (seconds) for training (total) and testing (averaged per scenario.) Training times are only reported for learning-based controllers.}
\label{tab:results-computational-time}
\begin{tabular}{l|l|c|c|c|c|c|c}
\specialrule{.15em}{.075em}{.075em}
\multicolumn{2}{c|}{} & \multicolumn{2}{c|}{TOU} & \multicolumn{2}{c|}{RTP} & \multicolumn{2}{c}{PC}\\ \hline
Controller & Variant & Train & Test & Train & Test & Train & Test \\
\specialrule{.15em}{.075em}{.075em}
OPT 
    & -- & -- & 25.5 & -- & 15.6 & -- & 34.6 \\ \hline
RBC
    & -- & -- & 1.3e-2 & -- & 1.3e-2 & -- & 1.3e-2 \\\hline
DPC 
    & -- & 3890.5 & 0.6 & 4166.9 & 0.8 & 3922.3 & 0.8 \\\hline
RLC 
    & -- & 14400.0 & 1.4 & 14400.0 & 1.4 & 14400.0 & 1.4 \\\hline
MPC (4 hr) 
    & MPC & -- & 434.0 & -- & 478.2 & -- & 460.5 \\
    & MPC-C & -- & 177.1 & -- & 175.0 & -- & 230.3 \\\hline
MPC (3 hr) 
    & MPC & -- & 312.7 & -- & 338.9 & -- & 330.8 \\
    & MPC-C & -- & 135.6 & -- & 134.3 & -- & 177.8  \\\hline
MPC (2 hr) 
    & MPC & -- & 190.9 & -- & 198.7 & -- & 202.5  \\
    & MPC-C & -- & 98.8  & -- & 97.8  & -- & 125.4  \\\hline
MPC (1 hr) 
    & MPC & -- & 90.0  & -- & 89.0  & -- & 94.4 \\
    & MPC-C & -- & 61.4  & -- & 61.6  & -- & 74.4 \\
    & MPC-CL & 21365.7 & 62.3  & 21988.5 & 62.3  & 25224.9 & 77.3 \\\hline
MPC (30 min) 
    & MPC & -- & 51.5  & -- & 51.1  & -- & 51.0 \\
    & MPC-C & -- & 47.7  & -- & 47.7  & -- & 53.5 \\
    & MPC-CL & 14740.3 & 48.6  & 14779.9 & 47.8  & 17137.9 & 55.0 \\\hline
MPC (15 min) 
    & MPC & -- & 30.7  & -- & 30.6  & -- & 29.6 \\
    & MPC-C & -- & 40.4  & -- & 40.5  & -- & 44.4 \\
    & MPC-CL & 11986.7 & 40.7  & 12037.7 & 40.7  & 13158.6 & 44.9 \\
\specialrule{.15em}{.075em}{.075em}
\end{tabular}
\end{table*}
\end{small}

The training process for all learning-based controllers is shown by the learning curves in Figure \ref{fig-lc-comparison}, which shows how the average episodic cost defined by \eqref{eqn:ocp-cost} changes with training episode (one episode corresponds to one simulated day.) For all curves the cost decreases as training progresses, indicating effective learning. For training RLC and DPC, we experimented with two learning rates: 1e-3 and 5e-5. With the smaller learning rate, both RLC and DPC converge along similar trajectories. With the larger learning rate, DPC learns faster and converges to a similar cost than with the smaller learning rate whereas RLC learns slower and converges to a higher cost. This difference occurs because at each training step, DPC calculates exact gradients using the building model whereas RLC estimates gradients from samples. Typically with more accurate gradient estimates, as in DPC, larger steps can be taken, speeding up training \cite[Section 2]{mccandlish2018empirical}. In contrast, with noisy estimates, using too large a learning rate, as shown for RL, leads to slow and sub-optimal convergence. Going forward we just consider DPC and RLC with the best learning rates (DPC-1e-3 and RLC-5e-5.)

Figure \ref{fig-lc-comparison} shows that MPC-CL has the highest sample efficiency, meaning it requires the least training samples to converge, followed by DPC and then RLC. MPC-CL requires fewer samples because it learns a terminal cost, which is simpler and thus easier to learn than the full control policy that DPC and RLC learn. DPC requires fewer samples than RLC, because DPC calculates exact gradients using the building model and therefore takes better-informed updates of the policy parameters than RLC, which can only approximate the gradients. Figure \ref{fig-lc-comparison} shows an additional benefit of MPC-CL: at the start of training its cost is already lower than that of RLC and DPC. This occurs because MPC-CL determines the control action based largely on the building model through an MPC approach; the terminal cost only shifts this action. RLC and DPC take actions based entirely on a policy, which much be trained. Thus, even when training starts MPC-CL chooses reasonable actions whereas RLC and DPC take essentially random and thus often unreasonable actions.

\begin{figure}[h]
\centering
\includegraphics[width=0.9\linewidth]{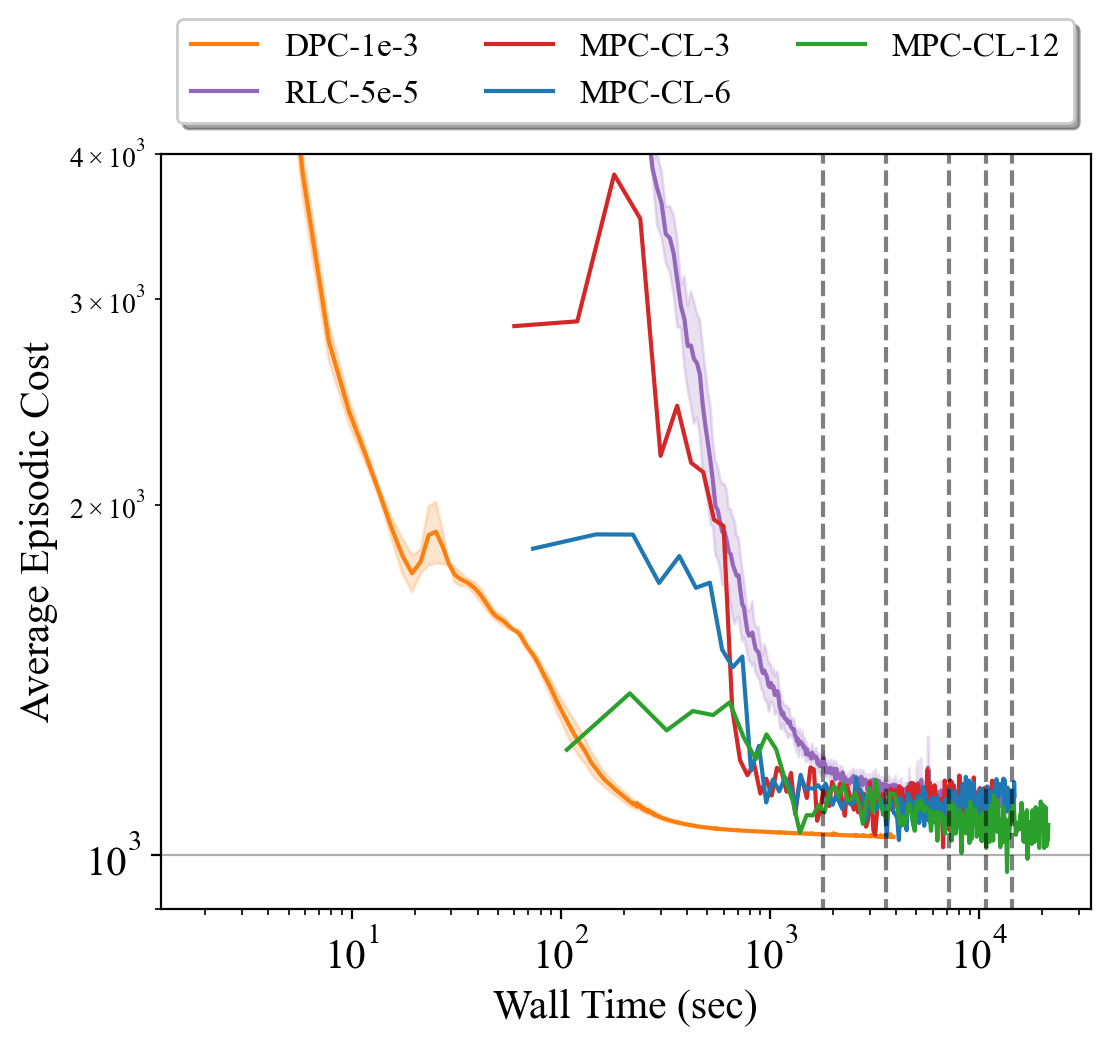}
\caption{Learning curves for training three learning-based building control algorithms under a TOU program. From left to right, the vertical dashed lines mark 30 mins, 1 hr, 2 hrs, 3 hrs and 4 hrs.}
\label{fig-lc-comparison-time}
\end{figure}

While MPC-CL has the highest sample efficiency, it doesn't have the highest computational efficiency, meaning it doesn't require the least amount of time for training; DPC does. This can be seen by comparing Figure \ref{fig-lc-comparison}, which uses training episode on the x-axis, and Figure \ref{fig-lc-comparison-time}, which uses training wall time. When using wall time, the learning curves for MPC-CL with different lookahead horizons no longer cluster; MPC-CL requires more training time when using a larger lookahead. Even the fastest training MPC-CL (with lookahead 12) has higher training time than DPC. MPC-CL performs worse on computational efficiency than sample efficiency, because it requires solving a MPC optimization problem at each step in the control episode during training, which is time-consuming; DPC and RLC just require a quick forward pass of a NN. Even though MPC-CL is more expensive for each sample than RLC, it still outperforms RLC on computational efficiency because it requires so many fewer samples than RLC does. The total training times are detailed in Table \ref{tab:results-computational-time}. 

\begin{figure*}[h]
\centering
\includegraphics[width=0.9\linewidth]{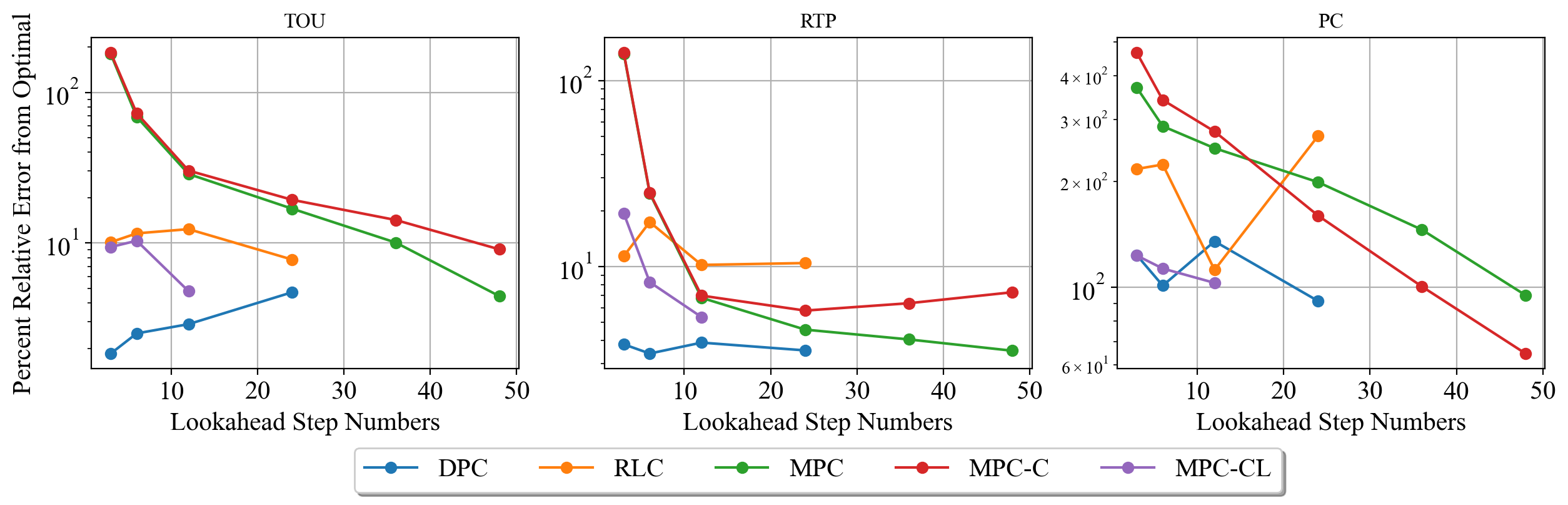}
\caption{Percent relative error from the optimal vs. lookahaead step numbers under three different DR programs. A lower relative error indicates the controller's performance is closer to that of OPT, meaning it provides better control performance.}
\label{fig:test-performance-compare}
\end{figure*}

\begin{figure*}[h]
\centering
\includegraphics[width=0.8\linewidth]{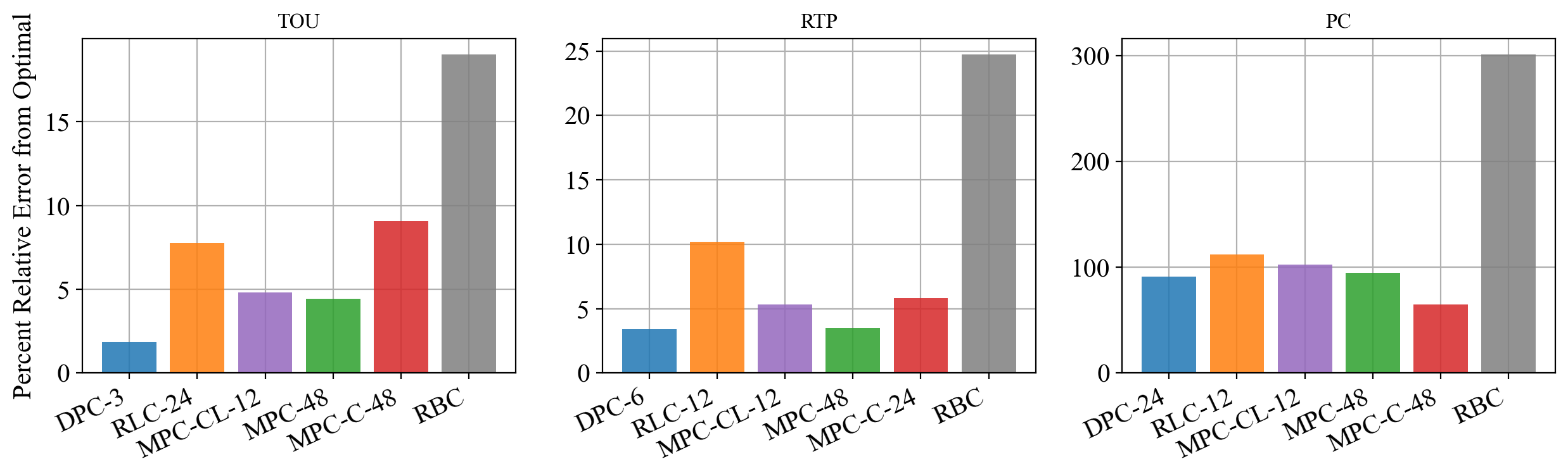}
\caption{Control performance comparison among the best-of-its-kind controllers of each type under three different DR programs.}
\label{fig:test-performance-compare-best-one}
\end{figure*}

\begin{figure*}[h]
\centering
\includegraphics[width=0.95\linewidth]{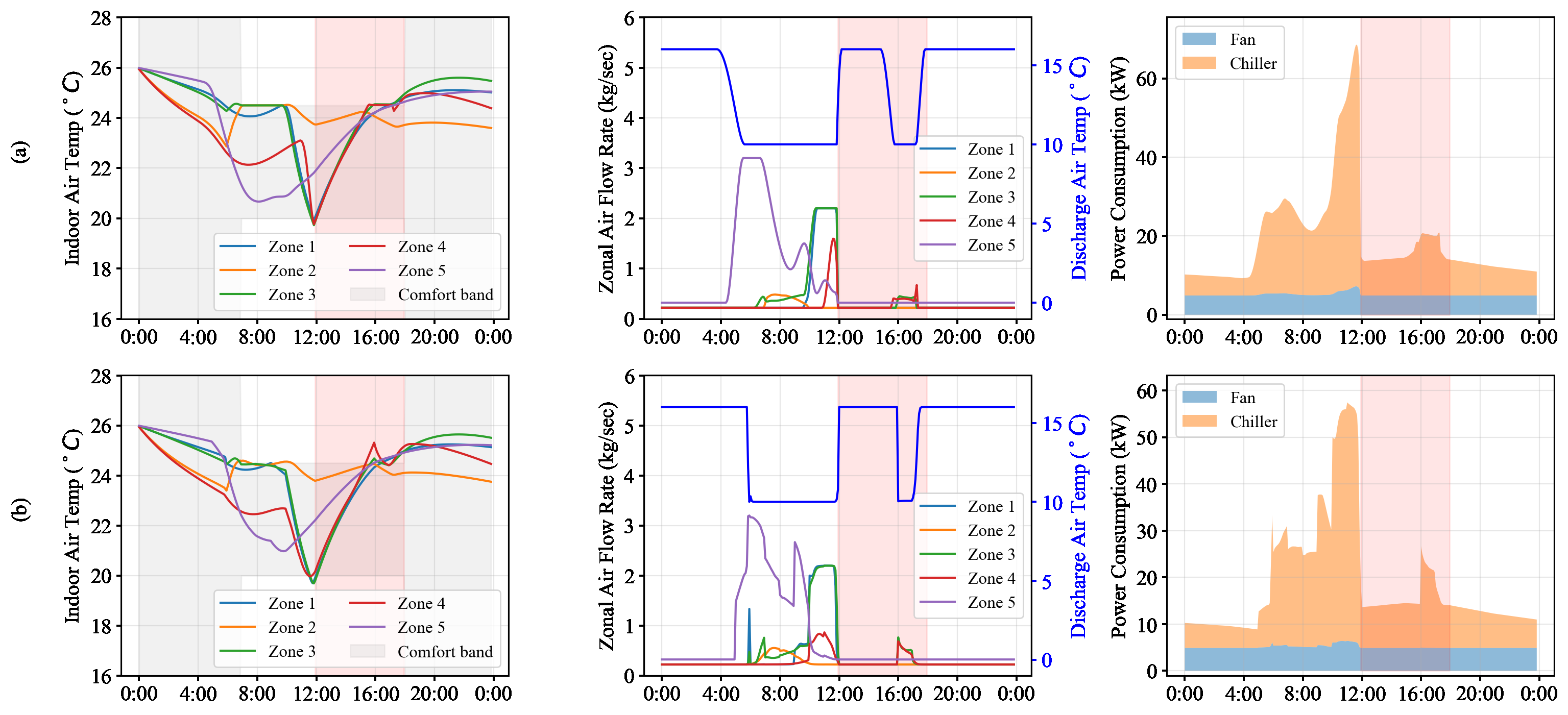}
\caption{Building operation history for a TOU program from OPT (a) and DPC-3 (b). The daily costs under these two controllers are 1344 and 1407 respectively. The pink shaded area represents the TOU peak price period, when the electricity price is 10x higher than during non-peak times. The left column shows the zone temperatures, the middle column shows the control signals on the primary y-axis and the discharge air temperature on the secondary y-axis, and the right column shows the power consumption by the fan and chiller.}
\label{fig:tou_specific_scenario}
\end{figure*}

\begin{figure*}[h]
\centering
\includegraphics[width=0.95\linewidth]{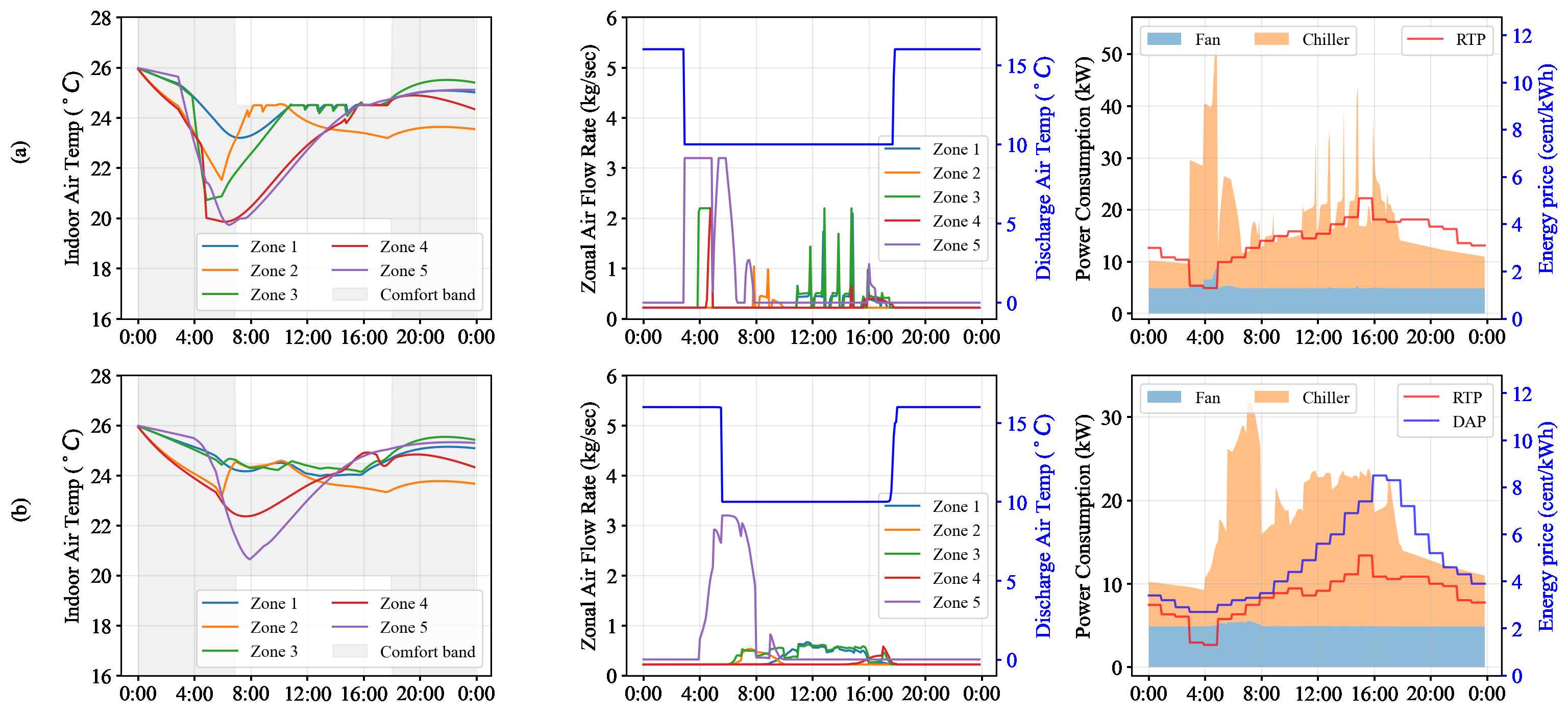}
\caption{Building operation history for a RTP program from OPT (a) and DPC-6 (b). Daily costs under these two controllers are 1378.5 and 1497.8 respectively. The left column shows the zone temperatures, the middle column shows the control signals on the primary y-axis and the discharge air temperature on the secondary y-axis, and the right column shows the power consumption by the fan and chiller.}
\label{fig:rtp_specific_scenario}
\end{figure*}

\subsection{Control Performance}\label{ssec:testing}

We compare the control performance of all methods first at a high level over all test days and then by a detailed examination for one specific test day. Through our detailed examination, we evaluate how well the learning-based controllers generalize to an out-of-distribution scenario.

\subsubsection{Overall Control Performance Comparison}

All controllers are tested for 31 days of test scenarios. For each controller, we average the daily control cost, \eqref{eqn:ocp-cost}, over all days and divide by the average daily control cost of the optimal control, OPT, presented in Section \ref{control:mpco}. The control percent error is shown in Figure \ref{fig:test-performance-compare} for TOU, RTP, and PC. 

For the MPC-type methods the control performance depends on lookahead length. In most cases, MPC, MPC-C, and MPC-CL demonstrate better performance if the lookahead horizon is longer. This is because we provide perfect forecasts for exogeneous inputs (as described in Assumption A.2 in Section \ref{subsec-problem-formulation}), so a longer lookahead allows the MPC-based controllers to plan further ahead. One exception is that in RTP, MPC-C performs slightly better with 2-hour lookahead horizon than with 2.5-hour and 3-hour. A possible explanation for this is that over a longer lookahead horizon, we accumulate errors from convexifying the building model and using DAP to predict RTP. Another result that requires additional consideration is that for the PC program MPC-C outperforms MPC for large lookaheads. We expect that this discrepancy comes from convexifying the problem: PC is a challenging case to solve since many of the constraints are binding and making the problem convex might make it easier to solve, leading to a better optimum. The difference might also come from the choice of solver itself.

By learning the terminal cost, MPC-CL can achieve the same control performance as a MPC controller that uses a longer lookahead horizon. For example, in TOU, MPC-CL with one-hour lookahead has a cost of 1071 which is close to that of MPC with four-hour lookahead, 1067, and is 19\% better than the cost of MPC with one-hour lookahead, 1315. Using a longer lookahead could further improve the control performance of MPC-CL; however, we don't consider lookaheads longer than one hour because the time and memory required for training becomes prohibitive. The training time grows large because MPC-CL does a backward pass through an optimization problem whose complexity scales with lookahead. The memory grows large because MPC-CL saves gradients through the building model for each sample in a batch. This problem does not occur for MPC and MPC-C because they do not use training and thus only need to solve an optimization problem and save the current computation. Lookahead length also does not pose a problem for DPC because it does a backward pass through and saves gradients for the building model once rather than repeatedly during an optimization problem.

For DPC and RLC, the lookahead horizon length $K$ affects the feature vector $\s(t)$ through the $K$-step forecast of exogeneous inputs. However, $K$ does not significantly affect the controller performance. This is because similar information is learned from other exogeneous inputs. Specifically, the controllers learn the relationship between time step and cost-saving control actions from $\mathbf{em}(t)$. For example, in TOU the controllers learn that around 11:00 am they should start pre-cooling to avoid using energy during peak hours, even though the peak hours start in 1-hour, beyond the 15-minute lookahaead. In this study we included lookahead in DPC and RLC to ensure a fair comparison with MPC-type methods that depend strongly on lookahead. However, due to the low impact of lookahead on DPC and RLC, future work can be simplified by omitting lookahead.

Figure \ref{fig:test-performance-compare-best-one} compares the performance of the best-of-its-kind controllers. For TOU, DPC with 15-min lookahead outperforms all other controllers and has only 1.8\% relative error from the optimal. For RTP, DPC with 30-min lookahead and MPC with 4-hr lookahead perform the best, with 3.4\% and 3.5\% relative error. For PC, MPC-C with 4-hr lookahead performs best, with 64.7\% relative error. The fact that three of the five methods show up as top performers indicates the broad range of methods that give good control performance. Beyond the top performers, MPC-CL and RLC also perform well, especially given their constraints. For TOU and RTP, MPC-CL outperforms MPC-C and performs nearly as well as MPC even though it uses a lookahead that is half to a quarter as long as that in MPC and MPC-C and uses a less accurate model than MPC --- both contributing to faster online control, as discussed in Section \ref{subsec-online-time}. As explained above, MPC-CL could perform better with a longer lookahead, but the time and memory for training then become prohibitive. Turning to RL, while it has the highest error of the controllers, it outperforms MPC and MPC-C for small lookahead and is the only method that does not directly use a building model.

Figure \ref{fig:test-performance-compare-best-one} reveals differences in the controllers' performances for PC relative to TOU and RTP. Whereas MPC outperforms MPC-C for TOU and RTP, the opposite is true for PC. As previously discussed, MPC-C might have outperformed MPC because convexifying the problem made it easier to solve. Another difference for PC is that all controllers' relative error is higher than for TOU and RTP. This occurred because we chose restrictive power limits to test the controllers under a challenging DR scenario, leading to constraint and power limit violations. The MPC-type methods would have required even longer lookahead horizons than those we considered to avoid these costs. For the learning-based methods, additional costs came from out-of-distribution days in the test data (days that are much hotter than any in the training data), which caused large constraint and power limit violations. We further discuss how the controllers generalize to out-of-distribution scenarios in Section \ref{subsec-controller-generalization}. 

\subsubsection{Detailed Examination of Controller Behavior}

In addition to comparing the controllers' overall control costs, here we examine their detailed behavior for one specific test day. We choose the hottest test day, which is the hardest day to satisfy thermal comfort constraints at a low cost. For conciseness, for each DR program, we compare only the method with best control performance and the OPT baseline.

Figure \ref{fig:tou_specific_scenario} shows the results under the TOU program for OPT and the best performing controller for this case, DPC with 15-min lookahead. Row (a) illustrates the indoor temperature, control set points, and power consumption over the selected test day under OPT. Row (b) gives the same information under DPC. DPC's control behavior is very close to that of OPT. Notably, both methods pre-cool starting around 10am so that the temperature is at the lower bound of the comfort band right before the TOU peak period starts; this minimizes cooling needed during TOU peak times without violating the comfort constraints before the peak period. In addition, both methods only significantly pre-cool the three of the five thermal zones that would otherwise require cooling during TOU peak times to avoid comfort violations.

Figure \ref{fig:rtp_specific_scenario} shows the results under RTP for OPT and DPC with 30-min lookahead. Similar to in the TOU case, OPT gives the optimal control actions, since there are no errors in the model or inputs, including the price (RTP). While, in contrast to OPT, DPC relies on approximate price information (DAP rather than RTP), DPC still shows reasonable control behavior. DPC pre-cools the largest zone, zone 5, early in the morning when the price is low. DPC slightly cools the other zones later in the day to satisfy the thermal constraints, similar to OPT.

\begin{figure*}
\centering
\includegraphics[width=0.95\linewidth]{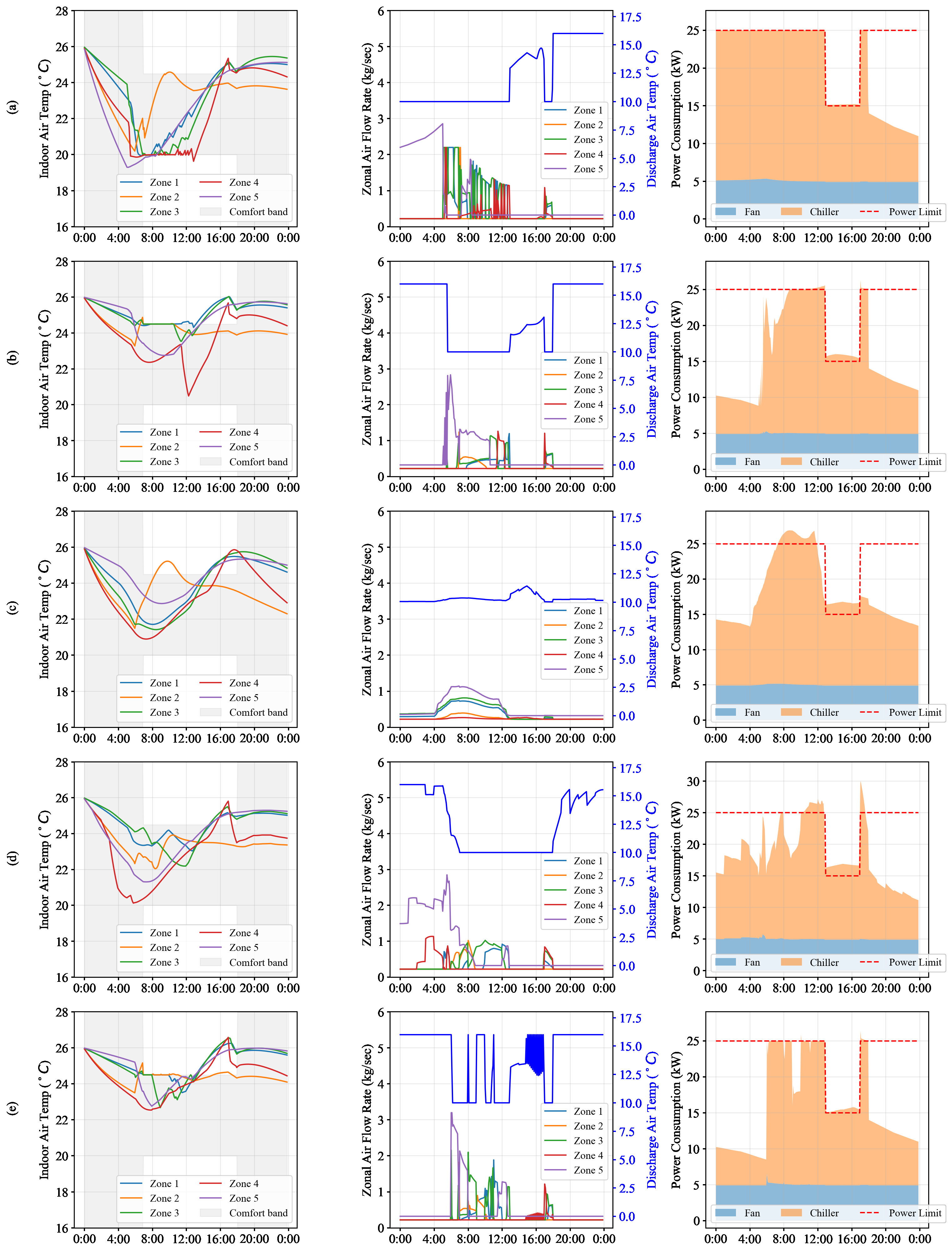}
\caption{Five controllers' performance under the power constrained DR program on the hottest day in testing scenarios: (a) OPT (control cost: 621.67), (b) MPC-C-48 (control cost: 2418.54) (c) DPC-24 (control cost: 3841.52) (d) RLC-12 (control cost: 4304.50) and (e) MPC-CL-12 (control cost: 3857.82). The left column shows the zone temperatures, the middle column shows the control signals on the primary y-axis and the discharge air temperature on the secondary y-axis, and the right column shows the power consumption by the fan and chiller.}
\label{fig:pc_specific_scenario}
\end{figure*}

Figure \ref{fig:pc_specific_scenario} rows (a) and (b) show the results under PC for OPT and MPC-C with 4-hour lookahead. Since the power limit is chosen to be restrictive and the test scenario is the hottest day in the test set, satisfying both the power and thermal comfort constraints is challenging. This is shown by the OPT control: the cooling starts at midnight and uses up all the allowed power before and during the DR event. For MPC-C, the 4-hour lookahead is not long enough to achieve this optimal control strategy; MPC-C only knows to start pre-cooling a couple hours before the DR event, which doesn't leave enough time to avoid power limit and thermal comfort constraint violations. Within the limitation of the lookahead, MPC-C does select reasonable control actions, pre-cooling to the extent it can. Note that this is the case with worst performance; on most days in the test set, MPC-C can satisfy both the power limit and thermal comfort constraints. The remaining rows in this figure are discussed in Section \ref{subsec-controller-generalization}.

\begin{figure}[h]
\centering
\includegraphics[width=0.85\linewidth]{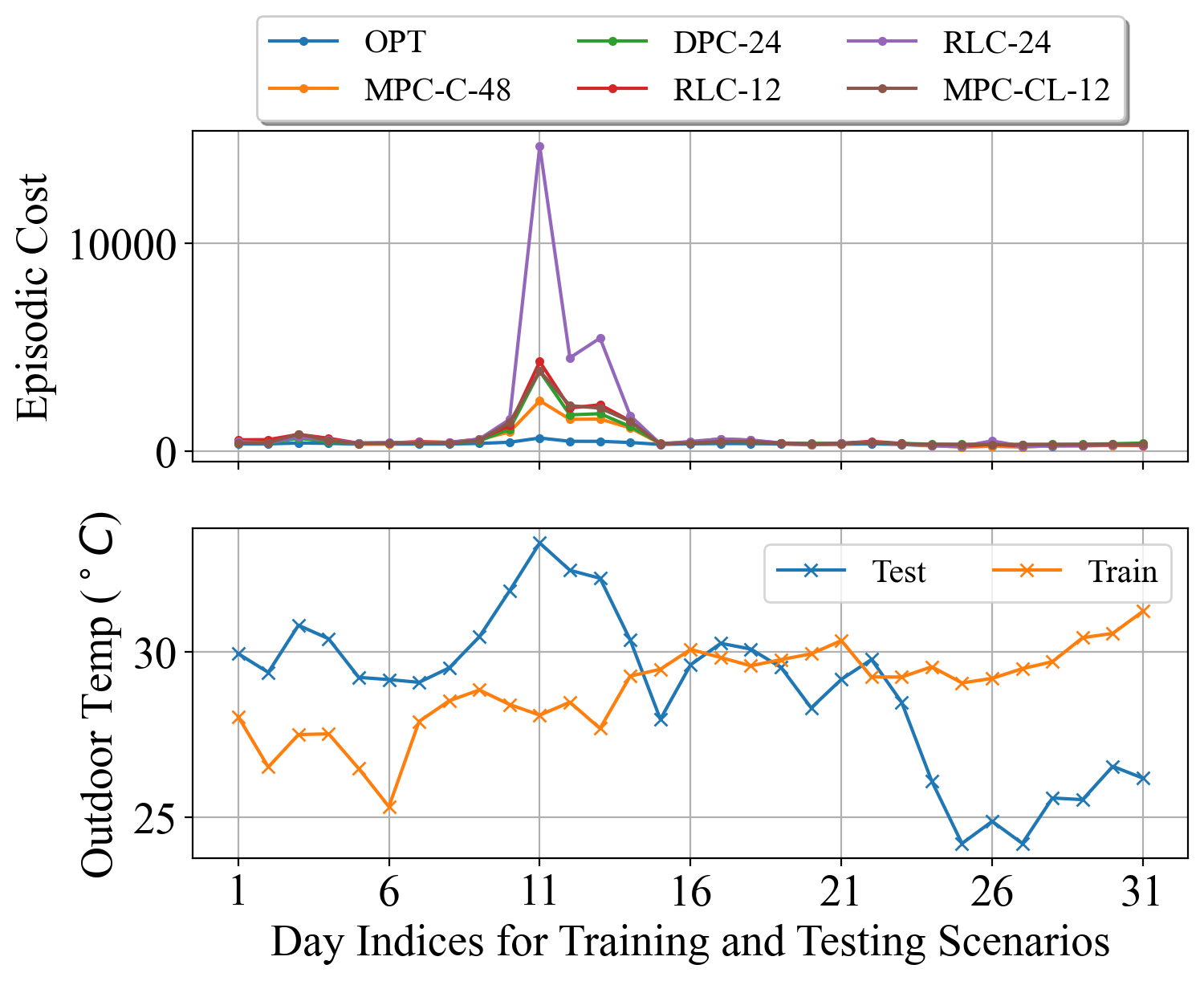}
\caption{Cost of six controllers under testing scenarios in the PC program (Top) and the daily average temperature for these scenarios (Bottom).}
\label{fig:cost-and-temp-power-constrained}
\end{figure}

\subsubsection{Controller Generalization}  \label{subsec-controller-generalization}

As described in Section \ref{ssec:training}, all learning-based controllers are trained to minimize the average cost over specific training scenarios. Therefore, the trained controllers should minimize the cost for scenarios that fall into the distribution of the training scenarios. Although our training and testing data come from two adjacent months and have similar distributions, there are some out-of-distribution days in the testing data, which have higher outside temperature than any day in the training data. The learning-based controllers generalize less well for these days. This is especially true for the PC case, because it is so challenging to satisfy for days with high temperature. This difficulty generalizing is evident in Figure \ref{fig:cost-and-temp-power-constrained}. The bottom row shows the daily average outdoor temperature for both months; the temperature for days 10-13 in the test data are higher than any in the training data. The top row shows that in those out-of-distribution test days, for PC, the learning-based methods (MPC-CL, RLC, DPC) have abnormally high costs --- even more so than the learning-free methods (OPT, MPC-C.)

Figure \ref{fig:pc_specific_scenario}(c)-(e) provides a more detailed view of control performance under PC for the hottest of the test days. This figure shows that DPC and RLC do take reasonable control actions, cooling all zones before and during the DR event, similar to OPT. However, they don't pre-cool sufficiently for this extremely hot day, leading to both power limit and thermal comfort constraint violations. This issue occurs not only for the model-free method, RL, but also for hybrid model- and learning-based methods, MPC-CL and DPC. Even though DPC is model-based, the model information is only directly used during training. During testing, control actions come from the trained policy without explicitly considering the building model. Therefore, DPC also struggles to generalize to out-of-distribution data. While, unlike DPC, in MPC-CL the model is used both during training and testing, MPC-CL still struggles to generalize. This is because in this out-of-distribution test case, the learned terminal cost no longer accurately reflects the cost beyond the lookahead horizon. Given the short lookahead of MPC-CL, the poor terminal cost estimate lead to thermal comfort violations. 

We designed this PC test case so that poor generalizability, such as that from out-of-distribution data, leads to high cost. However, our TOU and RTP cases show that this is not always true. For TOU and RTP the out-of-distribution data does not lead to high cost for the learning-based controllers. Additionally, not all out-of-distribution scenarios cause high cost for PC. For example, the outdoor temperature during testing days 25-27 is lower than for any training days but the costs remain low.

\subsubsection{Online Computational Time} \label{subsec-online-time}

\begin{figure}[h]
\centering
\includegraphics[width=0.9\linewidth]{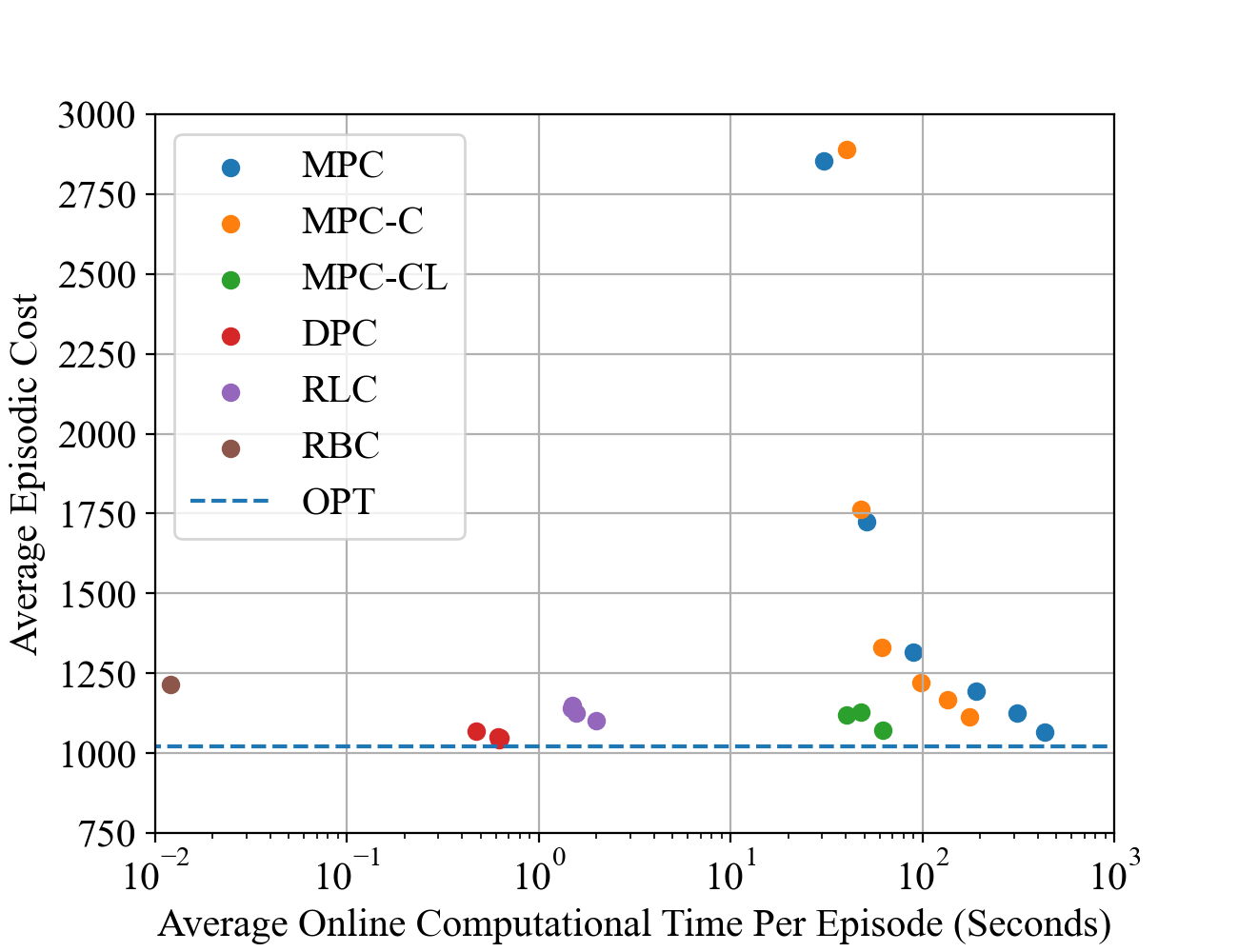}
\caption{Average control costs and average computational time per scenario for different controllers under TOU program. For MPC based controllers, different dots with same color represent different lookahaead length.}
\label{fig-TOU-cost-vs-time}
\end{figure}

To be used in real-time, controllers must generate actions quickly every step, meaning they must have low online computational time. This can be challenging for MPC-based methods, since they require solving a potentially computationally expensive optimization problem at every step. Solving this problem quickly can require more advanced computing hardware. Quick online computation is typically easier for policy-based methods, because they only require evaluating a predefined policy, such as simple rules for RBC and a NN for DPC and RL. This allows policy-based methods to be implemented on simple, low-cost hardware. Deciding whether it is worthwhile to use a method that requires more costly hardware comes down to a trade-off between the online computational time and the control performance; we show this comparison in Figure \ref{fig-TOU-cost-vs-time} for TOU. For this DR program, DPC shows the best balance between computational time and control performance, followed by RLC and MPC-CL. Though both only need policy network evaluation, which leads to about the same online computational time, RLC requires longer than DPC due to the overhead of the RL training framework in our implementation.

\section{Conclusion and Future Work}\label{sec-conclusion}

We compared a range of methods to control a multi-zone building under three demand response programs. This analysis was motivated by a surge in recent literature designing both model-based and learning-based building controllers. We briefly summarize the key takeaways of our study here.

\begin{itemize}
    \item Considering performance averaged across all the demand response programs and test days, the hybrid method differentiable predictive control (DPC) performed best. However, model-based methods performed better than DPC and other learning-based methods for challenging out-of-distribution test cases: unusually hot days under a power constrained (PC) program.
    \item Augmenting model predictive control (MPC) with a learnable terminal cost (MPC-CL) provides significant performance improvements over MPC with no terminal cost and is highly sample efficient. Despite its sample efficiency, MPC-CL takes significant training time since it requires solving an optimization problem for each sample and time step. In addition, its accuracy is limited by the training time and memory requirement of large lookaheads. MPC-CL is a novel method within the category of reinforced MPC (MPC-RL) which we first introduce in this paper and deserves further research to understand its opportunities and limitations.
    \item The final choice of controller depends on the DR program---as shown in this work, and may also be site specific---we considered only one site. Ultimately, the availability of training infrastructure (data, offline time, a potentially differentiable and convex model) and implementation infrastructure (online computing power) may be the biggest driver of the choice of controller.
\end{itemize}

In this study, we ran our experiments using a perfect model and exogeneous inputs. We chose this approach to enable a baseline comparison between model- and learning-based controllers. Future work should add error and use our results as a baseline to investigate that impact of error on each control method. Future work should also consider control of a higher-fidelity model~\cite{blum_boptest_2021} or a physical building. In addition, while we compared those control approaches that we consider to be the most promising, other approaches or variations on the approaches taken here could be compared using our released modular code~\cite{lbc_code}. Finally, though building control is used as an example in this paper, we believe the findings can be generalized to other applications

\section*{Acknowledgement}

This work was authored in part by the National Renewable Energy Laboratory, operated by Alliance for Sustainable Energy, LLC, for the U.S. Department of Energy (DOE) under Contract No. DE-AC36-08GO28308. This work was supported by the Laboratory Directed Research and Development (LDRD) Program at NREL via the Autonomous Urbanization: Mobility and Communities (AUMC) project. The views expressed in the article do not necessarily represent the views of the DOE or the U.S. Government. The U.S. Government retains and the publisher, by accepting the article for publication, acknowledges that the U.S. Government retains a nonexclusive, paid-up, irrevocable, worldwide license to publish or reproduce the published form of this work, or allow others to do so, for U.S. Government purposes.

This research was performed using computational resources sponsored by the Department of Energy's Office of Energy Efficiency and Renewable Energy and located at the National Renewable Energy Laboratory.

This research was supported in part by the U.S. Department of Energy Computational Science Graduate Fellowship under grant DE-SC0019323. Christiane Adcock is supported in part by a graduate fellowship award from Knight-Hennessy Scholars at Stanford University.

The first author received support from Maplewell Energy (Broomfield, CO, U.S.A.) for final revisions of this work.

The authors would also like to thank Xin Jin, Rohit Chintala, Christopher Bay, and Anna A. Moore for their insight and support.

\bibliography{reference}

\begin{thebibliography}{10}
\expandafter\ifx\csname url\endcsname\relax
  \def\url#1{\texttt{#1}}\fi
\expandafter\ifx\csname urlprefix\endcsname\relax\def\urlprefix{URL }\fi
\expandafter\ifx\csname href\endcsname\relax
  \def\href#1#2{#2} \def\path#1{#1}\fi

\bibitem{stott2016climate}
P.~Stott, How climate change affects extreme weather events, Science 352~(6293)
  (2016) 1517--1518.

\bibitem{room2021fact}
{White House Briefing Room}, Fact sheet: President biden sets 2030 greenhouse
  gas pollution reduction target aimed at creating good-paying union jobs and
  securing us leadership on clean energy technologies, The White House.

\bibitem{downie2021getting}
E.~Downie, Getting to 30-60: How china’s biggest coal power, cement, and
  steel corporations are responding to national decarbonization pledges, Tech.
  rep., Center on Global Energy Policy, Columbia University (2021).

\bibitem{somasundaram2014reference}
S.~Somasundaram, R.~Pratt, B.~Akyol, N.~Fernandez, N.~Foster, S.~Katipamula,
  E.~Mayhorn, A.~Somani, A.~Steckley, Z.~Taylor, {Reference guide for a
  transaction-based building controls framework}, Tech. rep., Pacific Northwest
  National Laboratory (2014).

\bibitem{leung2018decarbonizing}
J.~Leung, Decarbonizing us buildings, Tech. rep., Center for Climate and Energy
  Solutions (2018).

\bibitem{satchwell2021national}
A.~Satchwell, M.~A. Piette, A.~Khandekar, J.~Granderson, N.~M. Frick,
  R.~Hledik, A.~Faruqui, L.~Lam, S.~Ross, J.~Cohen, et~al., A national roadmap
  for grid-interactive efficient buildings, Tech. rep., Lawrence Berkeley
  National Lab (LBNL), Berkeley, CA (United States) (2021).

\bibitem{privara2011model}
S.~Privara, J.~{\v{S}}irok{\`y}, L.~Ferkl, J.~Cigler, Model predictive control
  of a building heating system: The first experience, Energy and Buildings
  43~(2-3) (2011) 564--572.

\bibitem{kim2020model}
D.~Kim, J.~E. Braun, Model predictive control for supervising multiple rooftop
  unit economizers to fully leverage free cooling energy resource, Applied
  Energy 275 (2020) 115324.

\bibitem{drgona2020all}
J.~Drgo{\v{n}}a, J.~Arroyo, I.~C. Figueroa, D.~Blum, K.~Arendt, D.~Kim, E.~P.
  Oll{\'e}, J.~Oravec, M.~Wetter, D.~L. Vrabie, et~al., {All you need to know
  about model predictive control for buildings}, Annual Reviews in Control 50
  (2020) 190--232.

\bibitem{tang2019model}
R.~Tang, S.~Wang, Model predictive control for thermal energy storage and
  thermal comfort optimization of building demand response in smart grids,
  Applied Energy 242 (2019) 873--882.

\bibitem{hu2019price}
M.~Hu, F.~Xiao, J.~B. J{\o}rgensen, R.~Li, Price-responsive model predictive
  control of floor heating systems for demand response using building thermal
  mass, Applied Thermal Engineering 153 (2019) 316--329.

\bibitem{wei2017deep}
T.~Wei, Y.~Wang, Q.~Zhu, {Deep reinforcement learning for building HVAC
  control}, in: Proceedings of the 54th Annual Design Automation Conference,
  2017, pp. 1--6.

\bibitem{zhang2019whole}
Z.~Zhang, A.~Chong, Y.~Pan, C.~Zhang, K.~P. Lam, {Whole building energy model
  for HVAC optimal control: A practical framework based on deep reinforcement
  learning}, Energy and Buildings 199 (2019) 472--490.

\bibitem{zhang2020edge}
X.~Zhang, D.~Biagioni, M.~Cai, P.~Graf, S.~Rahman, {An edge-cloud integrated
  solution for buildings demand response using reinforcement learning}, IEEE
  Transactions on Smart Grid 12~(1) (2021) 420--431.

\bibitem{zhang2022two}
X.~Zhang, Y.~Chen, A.~Bernstein, R.~Chintala, P.~Graf, X.~Jin, D.~Biagioni,
  {Two-Stage Reinforcement Learning Policy Search for Grid-Interactive Building
  Control}, IEEE Transactions on Smart Grid 13~(3) (2022) 1976--1987.

\bibitem{schmidt2018}
M.~Schmidt, C.~\r{A}hludn, Smart buildings as cyber-physical systems:
  data-driven predictive control strategies for energy efficiency, Renewable
  and sustainable energy reviews 90 (2018) 742--756.

\bibitem{wang2020}
Z.~Wang, T.~Hong, Reinforcement learning for building controls: The
  opportunities and challenges, Applied energy 260.

\bibitem{vazquez2019reinforcement}
J.~R. V{\'a}zquez-Canteli, Z.~Nagy, Reinforcement learning for demand response:
  A review of algorithms and modeling techniques, Applied energy 235 (2019)
  1072--1089.

\bibitem{xu2020one}
S.~Xu, Y.~Wang, Y.~Wang, Z.~O'Neill, Q.~Zhu, One for many: Transfer learning
  for building hvac control, in: Proceedings of the 7th ACM international
  conference on systems for energy-efficient buildings, cities, and
  transportation, 2020, pp. 230--239.

\bibitem{zhang2020transferable}
X.~Zhang, X.~Jin, C.~Tripp, D.~J. Biagioni, P.~Graf, H.~Jiang, Transferable
  reinforcement learning for smart homes, in: Proceedings of the 1st
  International Workshop on Reinforcement Learning for Energy Management in
  Buildings \& Cities, 2020, pp. 43--47.

\bibitem{pinto2022transfer}
G.~Pinto, Z.~Wang, A.~Roy, T.~Hong, A.~Capozzoli, Transfer learning for smart
  buildings: A critical review of algorithms, applications, and future
  perspectives, Advances in Applied Energy (2022) 100084.

\bibitem{yu2021district}
P.~Yu, H.~Hui, H.~Zhang, G.~Chen, Y.~Song, District cooling system control for
  providing operating reserve based on safe deep reinforcement learning, arXiv
  preprint arXiv:2112.10949.

\bibitem{kowli2012coordinating}
A.~Kowli, E.~Mayhorn, K.~Kalsi, S.~P. Meyn, {Coordinating dispatch of
  distributed energy resources with model predictive control and Q-learning},
  Tech. rep., Coordinated Science Laboratory, University of Illinois at
  Urbana-Champaign (2012).

\bibitem{arroyo2022reinforced}
J.~Arroyo, C.~Manna, F.~Spiessens, L.~Helsen, {Reinforced model predictive
  control (RL-MPC) for building energy management}, Applied Energy 309 (2022)
  118346.

\bibitem{drgona2018approximate}
J.~Drgo{\v{n}}a, D.~Picard, M.~Kvasnica, L.~Helsen, Approximate model
  predictive building control via machine learning, Applied Energy 218 (2018)
  199--216.

\bibitem{karg2020efficient}
B.~Karg, S.~Lucia, Efficient representation and approximation of model
  predictive control laws via deep learning, IEEE Transactions on Cybernetics
  50~(9) (2020) 3866--3878.

\bibitem{drgona2020learning}
J.~Drgona, A.~Tuor, D.~Vrabie, Learning stable adaptive explicit differentiable
  predictive control for unknown linear systems, arXiv preprint
  arXiv:2004.11184.

\bibitem{jin2020pontryagin}
W.~Jin, Z.~Wang, Z.~Yang, S.~Mou, Pontryagin differentiable programming: An
  end-to-end learning and control framework, Advances in Neural Information
  Processing Systems 33 (2020) 7979--7992.

\bibitem{amos2018differentiable}
B.~Amos, I.~Jimenez, J.~Sacks, B.~Boots, J.~Z. Kolter, Differentiable {MPC} for
  end-to-end planning and control, Advances in neural information processing
  systems 31.

\bibitem{lbc_code}
X.~Zhang, D.~Biagioni, J.~King, C.~Adcock, {U.S. DOE Laboratory Directed
  Research and Development}, {LBC (Learning Building Control)}, {Available}:
  \url{https://github.com/NREL/learning-building-control} (4 2022).
\newblock \href {http://dx.doi.org/10.11578/dc.20220418.3}
  {\path{doi:10.11578/dc.20220418.3}}.

\bibitem{energyplus}
Energyplus engineering reference, Tech. rep., National Renewable Energy
  Laboratory (2021).

\bibitem{chintala2015automated}
R.~H. Chintala, B.~P. Rasmussen, Automated multi-zone linear parametric black
  box modeling approach for building hvac systems, in: Dynamic Systems and
  Control Conference, Vol. 57250, American Society of Mechanical Engineers,
  2015, p. V002T29A004.

\bibitem{lazos_forecasting_2014}
D.~Lazos, A.~B. Sproul, M.~Kay, Optimisation of energy management in commercial
  buildings with weather forecasting inputs: {A} review, Renewable and
  Sustainable Energy Reviews 39 (2014) 587--603.

\bibitem{wachter2006}
A.~W{\"a}chter, L.~T. Biegler, On the implementation of a primal-dual interior
  point filter line search algorithm for large-scale nonlinear programming,
  Mathematical Programming.

\bibitem{bertsekas_4thed_2012}
D.~Bertsekas, Dynamic {Programming} and {Optimal} {Control}: {Volume} {I}, 4th
  Edition, Athena Scientific, 2012, google-Books-ID: qVBEEAAAQBAJ.

\bibitem{agrawal2019differentiable}
A.~Agrawal, B.~Amos, S.~Barratt, S.~Boyd, S.~Diamond, J.~Z. Kolter,
  Differentiable convex optimization layers, Advances in neural information
  processing systems 32.

\bibitem{schulman2017proximal}
J.~Schulman, F.~Wolski, P.~Dhariwal, A.~Radford, O.~Klimov, Proximal policy
  optimization algorithms, arXiv preprint arXiv:1707.06347.

\bibitem{drgona_DPC_2021}
J.~Drgoňa, A.~Tuor, E.~Skomski, S.~Vasisht, D.~Vrabie, {Deep Learning Explicit
  Differentiable Predictive Control Laws for Buildings}, IFAC-PapersOnLine
  54~(6) (2021) 14--19.

\bibitem{kingma2014adam}
D.~P. Kingma, J.~Ba, Adam: A method for stochastic optimization, arXiv preprint
  arXiv:1412.6980.

\bibitem{schulman2018}
J.~Schulman, P.~Moritz, S.~Levine, M.~Jordan, P.~Abbeel, High-dimensional
  continuous control using generalized advantage estimation, arXiv preprint
  arXiv::1506.02438v6.

\bibitem{liang2018rllib}
E.~Liang, R.~Liaw, R.~Nishihara, P.~Moritz, R.~Fox, K.~Goldberg, J.~Gonzalez,
  M.~Jordan, I.~Stoica, Rllib: Abstractions for distributed reinforcement
  learning, in: International Conference on Machine Learning, PMLR, 2018, pp.
  3053--3062.

\bibitem{buildingmodel}
M.~Deru, K.~Field, D.~Studer, K.~Benne, B.~Griffith, P.~Torcellini, B.~Liu,
  M.~Halverson, D.~Winiarski, M.~Rosenberg, M.~Yazdanian, J.~Huang, D.~Crawley,
  {U.S. Department of Energy commercial reference building models of the
  national building stock} (2011).

\bibitem{sdge2022tou}
{San Diego Gas \& Electric}, {Time-of-Use Plans}, {A}ccessed: Feb. 6th, 2022.
  [Online]. Available:
  \url{https://www.sdge.com/residential/pricing-plans/about-our-pricing-plans/whenmatters}
  (2022).

\bibitem{brandi2022comparison}
S.~Brandi, M.~Fiorentini, A.~Capozzoli, Comparison of online and offline deep
  reinforcement learning with model predictive control for thermal energy
  management, Automation in Construction 135 (2022) 104128.

\bibitem{comed2022rtp}
{The Commonwealth Edison Company}, {Real-time Hourly Prices}, {A}ccessed: Feb.
  6th, 2022. [Online]. Available:
  \url{https://hourlypricing.comed.com/live-prices/} (2022).

\bibitem{zhang2019iot}
X.~Zhang, M.~Pipattanasomporn, T.~Chen, S.~Rahman, An iot-based thermal model
  learning framework for smart buildings, IEEE Internet of Things Journal 7~(1)
  (2019) 518--527.

\bibitem{mccandlish2018empirical}
S.~McCandlish, J.~Kaplan, D.~Amodei, O.~D. Team, An empirical model of
  large-batch training, arXiv preprint arXiv:1812.06162.

\bibitem{blum_boptest_2021}
D.~Blum, J.~Arroyo, S.~Huang, J.~Drgoňa, F.~Jorissen, H.~T. Walnum, Y.~Chen,
  K.~Benne, D.~Vrabie, M.~Wetter, L.~Helsen, Building optimization testing
  framework ({BOPTEST}) for simulation-based benchmarking of control strategies
  in buildings, Journal of Building Performance Simulation 14~(5).

\end{thebibliography}

\end{document}